\pdfoutput=1
\documentclass[conference]{IEEEtran} 

\usepackage{url}
\usepackage{amsmath, amsfonts, amsthm, amssymb}
\usepackage{graphicx}
\usepackage{amsfonts}

\usepackage{authblk}

\usepackage{caption}
\usepackage{listings}
\usepackage{color}
\usepackage{amsmath}
\usepackage{url}
\usepackage{rotating}

\usepackage{color, colortbl}
\definecolor{Gray}{gray}{0.85}

\newcommand{\killpunct}[1]{}

\newcommand\T{\rule{0pt}{2.6ex}}       
\newcommand\B{\rule[-1.2ex]{0pt}{0pt}} 

\DeclareCaptionFont{white}{ \color{white} }
\DeclareCaptionFormat{listing}{
  \colorbox[cmyk]{0.43, 0.35, 0.35,0.01 }{
    \parbox{\textwidth}{\hspace{15pt}#1#2#3}
  }
}
\begin{document}

\title{Predicting Domain Generation Algorithms \\ with Long Short-Term Memory Networks}

\author{\large Jonathan Woodbridge\thanks{corresponding author}}
\author{\large Hyrum S. Anderson}
\author{\large Anjum Ahuja}
\author{\large Daniel Grant}

\affil{ \texttt{\{jwoodbridge,hyrum,aahuja,dgrant\}@endgame.com} \\ Endgame, Inc.\\Arlington, VA 22201}

\maketitle

\begin{abstract}
Various families of malware use domain generation algorithms (DGAs) to generate a large number of pseudo-random domain names to connect to a command and control (C2) server. In order to block DGA C2 traffic, security organizations must first discover the algorithm by reverse engineering malware samples, then generate a list of domains for a given seed.  The domains are then either preregistered, sink-holed or published in a DNS blacklist. This process is not only tedious, but can be readily circumvented by malware authors. An alternative approach to stop malware from using DGAs is to intercept DNS queries on a network and predict whether domains are DGA generated. Much of the previous work in DGA detection is based on finding groupings of like domains and using their statistical properties to determine if they are DGA generated.  However, these techniques are run over large time windows and cannot be used for real-time detection and prevention.  In addition, many of these techniques also use contextual information such as passive DNS and aggregations of all NXDomains throughout a network.  Such requirements are not only costly to integrate, they may not be possible due to real-world constraints of many systems (such as endpoint detection).  An alternative to these systems is a much harder problem: detect DGA generation on a per domain basis with no information except for the domain name. Previous work to solve this harder problem exhibits poor performance and many of these systems rely heavily on manual creation of features; a time consuming process that can easily be circumvented by malware authors. This paper presents a DGA classifier that leverages long short-term memory (LSTM) networks for real-time prediction of DGAs without the need for contextual information or manually created features.  In addition, the presented technique can accurately perform multiclass classification giving the ability to attribute a DGA generated domain to a specific malware family.  The technique is extremely easy to implement using open source tools allowing the technique to be deployed in almost any setting. Results are significantly better than all state-of-the-art techniques, providing 0.9993 area under the receiver operating characteristic curve for binary classification and a micro-averaged F1 score of 0.9906. In other terms, the LSTM technique can provide a 90\% detection rate with a 1:10000 false positive (FP) rate---a twenty times FP improvement over the next best method. Experiments in this paper are run on open datasets and code snippets are provided to reproduce the results.

\end{abstract}


\section{Introduction}
\label{introduction}
Many malware families contain domain generation algorithms (DGAs) to make preemptive defenses difficult.  Domains are generated pseudo-randomly in bulk (hundreds to tens-of-thousands per day) by a malware sample.  The malware then attempts to connect to all or a portion of these generated domains in hopes of finding a command and control (C2) server from which it can update, upload gathered intelligence, or pursue other malicious activities.  The malicious actor only needs to register a small number of these domains to be successful.  However, all the domains must be sinkholed, registered, or blacklisted before they go into use in order to preemptively defeat such an attack.  This defense becomes increasingly difficult as the rate of dynamically generated domains increases. 

Authors in \cite{kuhrer2014paint} presented a thorough review of the efficacy of blacklists.  As a part of this review, authors analyzed both public and private blacklists for DGA coverage, (i.e., how many domains generated by DGAs were contained in blacklists).  Public blacklists were surprisingly lacking in terms of DGA coverage with less than 1.2\% of DGAs analyzed by the authors being contained in any of the blacklists.  Vendor provided blacklists fared better, but had mixed results over malware families with coverage varying from 0\% to 99.5\%.  These results suggest that blacklists are useful, but must be supplemented by other techniques to provide a more adequate level of protection.

Another approach to combating malware using DGAs is to build a DGA classifier.  This classifier can live in the network sniffing out DNS requests and looking for DGAs.  When DGAs are detected, the classifier notifies other automated tools or network administrators to further investigate the origin of a DGA. Previous work in DGA detection can be broken down into two categories: retrospective detection and real-time detection.  Retrospective detection makes bulk predictions on large sets of domains and are designed as a reactionary system that cannot be used for real-time detection and prevention \cite{antonakakis2012throw, yadav2010detecting, yadav2012detecting}.  In these systems, sets of domains are broken down into groupings using clustering with the intent to generate statistical properties of each grouping.  Classification is accomplished by generating templates during training and using statistical tests (e.g., Kullback-Leibler divergence) to classify groups of potential DGAs.  In addition, these techniques incorporate contextual information such as HTTP headers, NXDomains across a network, and passive DNS to further improve performance.  Much of the previous work in DGA detection falls in the former category and, unfortunately, does not meet the needs of many real-world security applications that require real-time detection and prevention \cite{krishnan2013crossing}. In addition, it is often unrealistic for many security applications to use contextual information.  For example, endpoint detection and response (EDR) systems run on endpoints and hosts and have strict performance requirements on processing, network, and memory usage.  Aggregating such contextual information from the network to each endpoint requires far too much overhead and is not practical for a real-world deployment.

Real-time detection techniques attempts to classify domains as DGA generated on a per domain basis using only the domains' names (i.e., no additional contextual information).  Real-time detection is a considerably harder problem than retrospective techniques and techniques often exhibit performance far too low for a real-world deployment.  (Suprisingly, authors in \cite{krishnan2013crossing} found that retropsective techniques had similarly bad performance!)  Many of the previous real-time approaches use hand picked features (e.g., entropy, string length, vowel to consonant ratio, etc.) that are fed into a machine learning model, such as a random forest classifier. Using hand-crafted features have two major drawbacks.  First, hand-crafted features are easy to circumvent.  Second, deriving hand-crafted features is a time consuming process.  If, and when, a malicious actor derives a new DGA family around beating a set of features, security professionals will need to spend considerable time creating new features.  To the best of our knowledge, authors in \cite{antonakakis2012throw} presented the first (and only until this paper) featureless real-time technique by using Hidden Markov Models (HMMs). However, as shown later in the paper, HMMs perform quite poorly on detecting DGAs.  To note, the HMMs in \cite{antonakakis2012throw} were part of a much larger retrospective detection system.


This paper presents a feature-less real-time technique using Long Short-Term Memory networks (LSTMs) to classify DGAs.  This technique has four significant advantages over other techniques in the literature.  First, the LSTM DGA classifier is featureless, in that it operates on raw domain names (e.g., google.com, facebook.com, etc.).  If a new family of DGA appears, then the classifier can be retrained without the tedious step of hand picking features.  LSTMs work largely as a black box making it very difficult for adversaries to reverse engineer and beat a classifier without the same training set.  Second, the presented technique has a significantly better true positive rate/false positive rate over previously published retrospective and real-time approaches.  Third, the technique also works in a multiclass classification setting.  Therefore, the algorithm not only provides a binary decision of whether a domain is DGA or not, but can accurately fingerprint a unique DGA's structure.  Fourth, the presented algorithm can classify in real-time using absolutely no contextual information.  Classification of a domain takes $20\,\mathrm{ms}$ on commodity hardware.\footnote{Apple MacBook Pro with a 2.2 GHz Intel Core i7 and 16GB of memory} The technique is trivial to implement and can run on virtually any security environment.  In fact, all the code required to implement this system is provided in this paper demonstrating its ease of deployment.

In this paper, we make the following contributions. We

\begin{enumerate}
  \item introduce an LSTM network to predict DGA generated domains, which to our knowledge, is the first application and in-depth analysis of deep learning to this domain;
  \item present complete experimental results showing significant improvements over previous techniques (both real-time and retrospective) in the literature using open datasets; and
  \item provide source code to reproduce results.
\end{enumerate}

To allow for easily reproducible results, Python source code built on the open source framework Keras \cite{chollet2016} is provided.  Experiments were run on GPU hardware, but it's possible to run all experiments on commodity desktop or laptop hardware.  An overview of LSTMs and previous work is discussed in Section \ref{background}. Details of reproducing the results are given in Sections \ref{method} and \ref{expsetup}.  Full results are given in Section \ref{results} with suggestions for future work in Section \ref{conclusion}.

\section{Background}
\label{background}
Domain fluxing is a technique used by botnets and command-and-control (C2) servers to create many domains using a Domain Generation Algorithm (DGA) \cite{knysz2011good, stone2011analysis}.  All botnets and C2 servers in the same infrastructure use the same seeded algorithm such that they all create the same pseudorandomly generated domains.  A subset of these domains are registered by the C2 servers while each botnet iterates through the DGA generated domains until it finds one that is registered.  To further complicate the process, C2 servers continually switch to new DGA generated domains making blacklist creation and take down efforts difficult.

One approach to combating domain fluxing is to reverse engineer a piece of malware and its respective DGA \cite{stone2011analysis}.  Once a DGA and its respective seed is known, future domains can be registered and used as an impostor C2 server to hijack botnets (a process known as sinkholing).  Once a campaign has been hijacked, adversaries must redeploy new botnets with updated seeds to continue.

Blacklisting is another approach to combat domain fluxing \cite{kuhrer2014paint}.  DGA generated domains are added to a blacklist that can be used by a network administrator to block connections to potential C2 servers.  However, both blacklists and sinkholing are only effective when both the algorithm and seed used by a campaign is known.

\subsection{Domain Generation Algorithms}
This paper evaluates the ability to classify DGA generated domains from 30 different types of malware.  Malware families include ransomware, such as \texttt{Cryptolocker} \cite{ward2014cryptolocker, cryptolocker} and \texttt{Cryptowall} \cite{hampton2015ransomware}, banking trojans, such as \texttt{Hesperbot} \cite{cherepanov2013hesperbot}, and general information-stealing tactics, such as \texttt{ramnit} \cite{w32ramnit}.

DGA techniques vary in complexity from simple uniformly generated domain names to those that attempt to model distributions that are seen in real domains.  \texttt{ramnit}, for example, creates domains with a series of divides, multiplies and modulos computed on a seed \cite{w32ramnit} while \texttt{suppobox} creates domains by concatenating two random strings (typically taken from the English language) \cite{suppoboxbh}.

Predicting DGA generated domains from such algorithms as \texttt{suppobox} is extremely difficult without using contextual information.  In fact, the LSTM technique presented in this paper was the only real-time technique able to classify such domains.  

\subsection{DGA Classification}
DGA classification can be a useful component of a domain reputation system.  Domain reputation systems have the task of assigning a trustworthy score of a domain.  This score typically varies from 0 (most benign) to 1 (most malicious).  Domain reputation systems typically incorporate many pieces of heterogeneous data, such as passive DNS (pDNS), to make decisions on a domain's reputation \cite{antonakakis2010building, bilge2011exposure, bilge2014exposure}.  DGA classification is one piece of information that can help assign a reputation to a domain. Previous approaches to DGA classification can be roughly broken down into two categories:

\begin{enumerate}
  \item \textit{Retrospective}: classifying domains in groups to take advantage of bulk statistical properties or common contextual information; and
  \item \textit{Real-time}: classifying domains individually with no additional contextual information.
\end{enumerate}

Authors in \cite{yadav2010detecting, yadav2012detecting} detect DGAs by using both unigram and bigram statistics of domain clusters.  The training set is separated into two subsets: those generated by a DGA and those not generated by a DGA.  The distributions of both unigrams and bigrams are calculated for both the subsets.  Classification occurs in batches.  Each batch of unknown domains is clustered by shared second level domain and domains sharing the same IP address.  The unigram and bigram distributions are calculated for each cluster and compared to the two known (labeled) subsets using the Kullback-Leibler (KL) distance.  In addition, the authors use the Jaccard distance to compare bigrams between clusters and the known (labeled) sets as well.

Authors in \cite{antonakakis2012throw} apply a similar clustering process to classify domains with unsuccessful DNS resolutions.  To train, statistical features are calculated for each subset of labeled DGA generated domains, such as \texttt{Bobax}, \texttt{Torpig}, and \texttt{Conficker.C}. Unknown domains are clustered by statistical characteristics such as length, entropy, and character frequency distribution, as well as shared hosts requesting the domain (i.e., cluster two domains together if the same host made a DNS query for both domains).  Next, statistical features are calculated for each cluster and compared to the training subsets to classify the clusters as formed by a known DGA. If a cluster is classified as belonging to a known DGA, the host is deemed to be infected.

Once a host is deemed to be infected with a DGA-bot, the authors attempt to identify the bots active C2 server.  This stage of the process uses a Hidden Markov Model trained on each known family of DGA and applied to single domains (i.e., this technique follows the same assumptions as the LSTM technique proposed by this paper).  Each domain with a successful DNS request is fed through each HMM.  If a domain receives an adequate score (i.e., greater than some threshold $\theta$), the domain is labeled as a DGA.  The threshold is learned at training time and set to a maximum false positive rate of 1\%.  We use this HMM technique as one of our comparisons to previous work.  

The aforementioned techniques (with exception to the HMM technique in \cite{antonakakis2012throw}) are accomplished retrospectively.  Authors in \cite{krishnan2013crossing} perform an in-depth comparison of these techniques and discuss two important findings.  First, retrospective techniques are too slow for most real-world deployments and often take hours to detect malicious domains.  Second, the performance of these systems are quite poor in terms of false positives and true positives. These authors present their own technique that overlaps both retrospective and real-time techniques.  They apply an online form of sequential hypothesis testing to NXDomains only.  Clients in a network are given an evolving score based on the number and maliciousness of NXDomains.  A client can be labeled as malicious or benign once its score goes above or below predefined thresholds.  While this system is a big improvement over retrospective systems, it has three main drawbacks. First, detection is not always in real-time as a client takes time to build an appropriate score.  Authors reported that only $83\%$ of domains were detected in time to prevent a connection.  Second, performance of their system is considerably less than most real-time solutions as we show in section \ref{results}. Third, their system cannot perform multiclass classification as their system bases classification solely on the presence of NXDomains.

Authors in \cite{schiavoni2014phoenix} present a real-time DGA classifier that uses two basic linguistic features named \textit{meaningful characters ratio} and \textit{$n$-gram normality score}.  The \textit{meaningful characters ratio} calculates the ratio of characters in a domain that comprise of a meaningful word.  For example, \emph{facebook} has a ratio of 1 as all character in the domain are covered by the words \emph{face} and \emph{book} while \emph{face1234} has a ratio of 0.5 as only half of its character are covered by the word \emph{face}.  The \textit{$n$-gram normality score} is calculated by finding n-grams with $n \in {1, 2, 3}$ within a domain and calculating their count in the English language.   The mean and covariance of these four features are calculated from a benign set (\emph{Alexa} top 100,000).  Unknown domains are then classified by their Mahalanobis distance to the benign set (i.e. a larger distance is indicative of a DGA generated domain).  

The approach in \cite{schiavoni2014phoenix} is used as a filter step.  Once domains have been classified as a DGA they are fed to a clustering technique (similar to those described above) to further classify the domains.

Section \ref{results} shows a comparison of our technique to both retrospective and real-time systems.  Our technique significantly outperforms retrospective techniques and the comparison is brief and compares findings to those in \cite{krishnan2013crossing}. An in depth comparison is performed between our technique and the aforementioned real-time systems.  More specififcally, we compare our technique to the HMM defined by \cite{antonakakis2012throw} as well as a Random Forest Classifier trained on features defined in \cite{antonakakis2012throw, yadav2010detecting, yadav2012detecting, schiavoni2014phoenix}.  We do not perform an in depth comparison on the full systems as defined in \cite{antonakakis2012throw, yadav2010detecting, yadav2012detecting} as they are retrospective systems and have already been shown to perform far worse than our system \cite{krishnan2013crossing}.  
 
\subsection{LSTM Networks}
In a variety of natural language tasks, recurrent neural networks (RNNs) have been used to capture meaningful temporal relationships among tokens in a sequence \cite{robinson1994application,mikolov2010recurrent,graves2012sequence,bengio2013advances}. The key benefit of RNNs is that they incorporate contextual (state) information in their mapping from input to output.  That is, the output of a single RNN cell is a function of the input layer and previous RNN activations.  Due to long chains of operations that are introduced by including self-recurrent connections, the output of a traditional RNN may decay exponentially (or, more rarely but catastrophically explode) for a given input, leading to the well-known  \textit{vanishing gradients} problem.  This makes learning long-term dependencies in an RNN difficult to achieve.  

The problem of vanishing gradients is a key motivation behind the application of the Long Short-Term Memory (LSTM) cell \cite{hochreiter1997long, gers2000learning, gers2003learning}, which consists of a state that can be read, written or reset via a set of programmable gates.  The cell's state has a self-recurrent connection that allows the cell to exactly retain state between time steps.  However, that state may be modulated by a new input via an input gate, which effectively multiplies the input by a number that ranges between 0 and 1 (sigmoid activation) or -1 and 1 (tanh activation).  Likewise, a forget gate modulates the self-recurrent state connection by a number between 0 and 1.  Thus, if the input gate modulates the input with 0, and the forget gate modulates the recurrent connection with 1, the cell ignores the input and perfectly retains state.  On the other hand, a 1 (input) and a 0 (forget) causes the cell's state to be overwritten by the input.  And in the case of a 0 (input) and 0 (forget), the state is reset to 0.  Finally, an output gate modulates the contribution of the cell's state to the output, which propagates to the input gates of LSTM cells across the layer, as well as to subsequent layers of the network. 

The LSTM cell's design with multiplicative gates allows a network to store and access state over long sequences, thereby mitigating the vanishing gradients problem.  For our use with domain names, the state space is intended to capture combinations of letters that are important to discriminating DGA domains from non-DGA domains.  This flexible architecture generalizes manual feature extraction via bigrams, for example, but instead learns dependencies of one or multiple characters, whether in succession or with arbitrary separation.


\section{Method}
\label{method}
We employ an LSTM network for detecting DGAs.  The model has the following advantages:
\begin{itemize}
\item the model accepts variable-length character sequences as input, so that there is no auxiliary requirement for feature extraction\footnote{In experiments, we employ a trivial pre-processing step to remove top-level domains and convert all characters to lowercase.};
\item the model is very compact, comprised simply of an embedding layer, an LSTM network layer, and a fully connected output layer that is simple logistic (or for multiclass, multinomial logistic) regression; and
\item although training on a large dataset is computationally intensive, the shallow structure allows for very fast query times.
\end{itemize}

\begin{figure}
\centering
\includegraphics[scale=0.5]{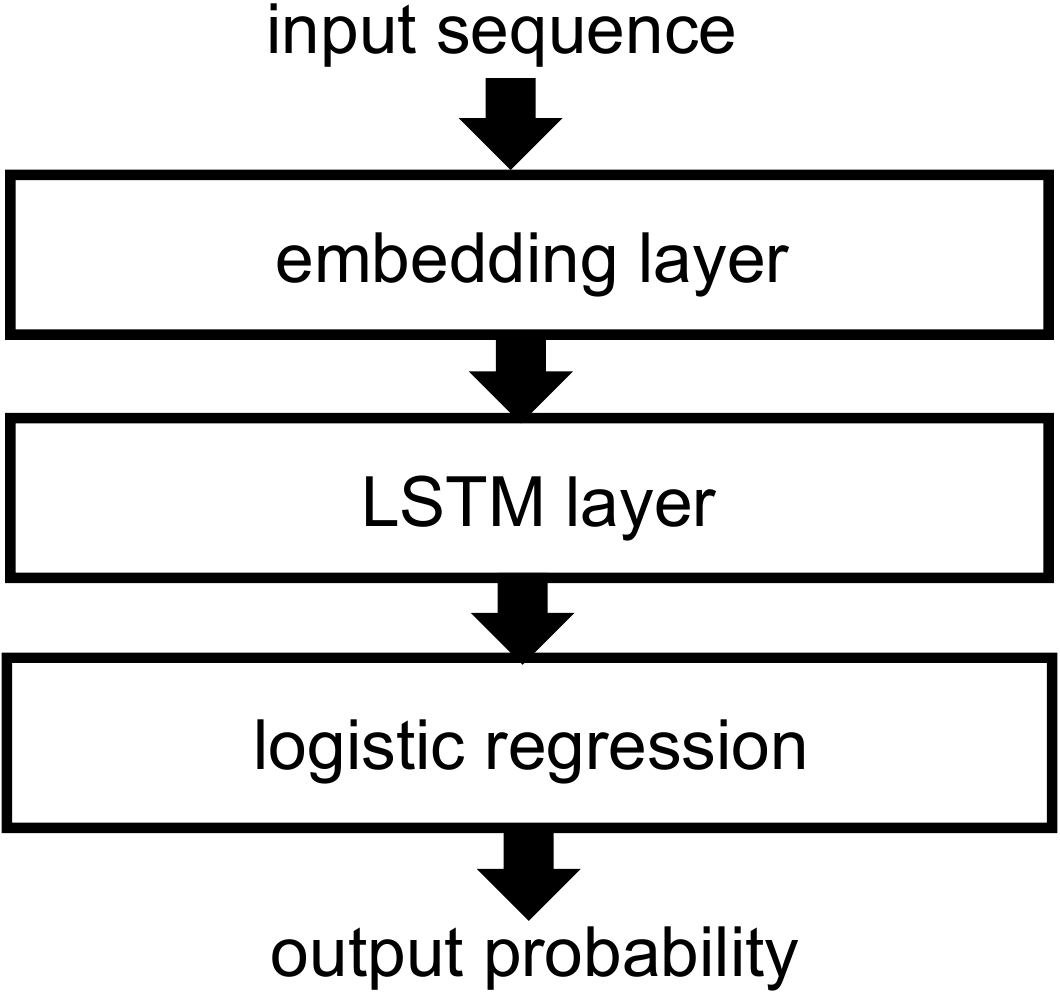}
\caption{Our model consists of an embedding layer, an LSTM layer that serves essentially as a feature extractor, and a logistic regression classifier.} \label{fig:model}
\end{figure}

A graphical depiction of our model is shown in Fig. \ref{fig:model}.  To prevent overfitting when training neural networks, it is common practice to employ dropout. Dropout consists of randomly removing a random subset of edges between layers of a network during each iteration of training, but restoring their contribution at test time.  We apply dropout after the LSTM layer prior to logistic regression.

The embedding layer projects $\ell$-length sequences of input characters from the input domain $\mathcal{S} \subset \mathcal{Z}^{\ell}$ to a sequence of vectors $\mathcal{R}^{d\times \ell}$, where $\ell$ is an upper bounded length determined from the training set.  The input domain consists of non-redundant valid domain name characters (lowercase alphanumeric, period, dash and underscore), and the output dimension $d$ is a tunable parameter that represents an embedding.  In our model, we choose $d=128>|\mathcal{S}|$ to provide additional degrees of freedom to the model, but preliminary experiments showed that results are relatively insensitive to the particular choice of $d$.

The LSTM layer can be thought of as implicit feature extraction, as opposed to explicit feature extraction (e.g., \textit{n}-grams) used in other approaches.  Rather than represent domain names explicitly as a bag of bigrams, for example, the LSTM learns patterns of characters (or in our case, embedded vectors) that maximize the performance of the second classification layer. In our experiments we compare the LSTM model to an explicit bigram logistic regression model.

All LSTM code was written in Python using the Keras framework \cite{chollet2016}.  Two models are generated: one for a binary classification and one for a multiclass classification.  Code for the binary classification is shown in Fig. \ref{fig:code_binary} and the multiclass classification in Fig. \ref{fig:code_multi}.

The two code examples have a few small differences.  The final dense layer goes from an output of one value in the binary classifier (line 15) to \texttt{nb\_classes} in the multiclass classifier (line 17).  A binary decision only requires a single value from $[0,1]$ where 0 is the most benign and 1 is the most DGA.  The multiclass model produces \texttt{nb\_classes} scores, one for each family known by the classifier, where multinomial logistic regression is employed on \emph{softmax}ed activations on line 18 to encode a distribution that sums to unity.


\begin{figure}
\resizebox{\columnwidth}{!}{%
\begin{lstlisting}[language=Python]
from keras.preprocessing import pad_sequences
from keras.models import Sequential
from keras.layers.core import Dense
from keras.layers.core import Dropout
from keras.layers.core import Activation
from keras.layers.embeddings import Embedding
from keras.layers.recurrent import LSTM

model=Sequential()
model.add(Embedding(max_features, 
                    128, 
                    input_length=75))
model.add(LSTM(128))
model.add(Dropout(0.5))
model.add(Dense(1))
model.add(Activation('sigmoid'))

model.compile(loss='binary_crossentropy',
              optimizer='rmsprop')

# Pad sequence where sequences are case 
# insensitive characters encoded to 
# integers from 0 to number of valid 
# characters
X_train=sequence.pad_sequences(X_train, 
                               maxlen=75)

# Train where y_train is 0-1
model.fit(X_train, y_train, 
          batch_size=batch_size, nb_epoch=1)
\end{lstlisting}
}
\caption{Binary LSTM Code}
\label{fig:code_binary}
\end{figure}

\begin{figure}
\resizebox{\columnwidth}{!}{%
\begin{lstlisting}[language=Python]
from keras.preprocessing import pad_sequences
from keras.models import Sequential
from keras.layers.core import Dense
from keras.layers.core import Dropout
from keras.layers.core import Activation
from keras.layers.embeddings import Embedding
from keras.layers.recurrent import LSTM

model=Sequential()
model.add(Embedding(max_features, 
                    128, 
                    input_length=75))
model.add(LSTM(128))
model.add(Dropout(0.5))
# nb_classes is the number of classes in
# the training set
model.add(Dense(nb_classes)) 
model.add(Activation('softmax'))

model.compile(loss='categorical_crossentropy',
              optimizer='rmsprop')

# Pad sequence where sequences are case 
# insensitive characters encoded to 
# integers from 0 to number of valid 
# characters
X_train=sequence.pad_sequences(X_train, 
                               maxlen=75)

# Train where y_train is one-hot encoded for
# each class
model.fit(X_train, y_train, 
          batch_size=batch_size, nb_epoch=1)
\end{lstlisting}
}
\caption{Multiclass LSTM Code}
\label{fig:code_multi}
\end{figure}

\section{Experimental Setup}
\label{expsetup}
In the following section, we describe details of our experimental setup in evaluating DGA classifiers in a binary experiment (DGA vs. non-DGA) and multiclass experiment (which DGA?) using publically available domain names and DGA data.

\subsection{Evaluation Metrics}
Precision, Recall, $F_1$ score, and Receiver Operating Characteristic (ROC) are the four evaluation metrics used to compare the LSTM classification technique to other state-of-the-art techniques.  Precision is defined as

\begin{equation*}
\label{eq:precision}
\textrm{Precision} = \frac{\sum{\textrm{True Positive}}}{\sum{\textrm{True Positive}}+\sum{\textrm{False Positive}}},
\end{equation*}

\noindent
and measures the purity of all positively labeled instances (i.e., the ratio of correct positively labeled instances to all positively labeled instances).  Recall is defined as

\begin{equation*}
\label{eq:recall}
\textrm{Recall} = \frac{\sum{\textrm{True Positive}}}{\sum{\textrm{True Positive}}+\sum{\textrm{False Negative}}},
\end{equation*}

\noindent
and measures the completeness of positively labeled instances (i.e., the ratio of correct positively labeled instances to all instances that should have been labeled positive).  $F_1$ score is the harmonic mean of Precision and Recall: 

\begin{equation*}
\label{eq:f1}
F_1 = 2 \cdot \frac{\textrm{Precision} \cdot \textrm{Recall}}{\textrm{Precision}+\textrm{Recall}}.
\end{equation*}

ROC measures the trade-off of the true positive rate (TPR) to false positive rate (FPR) where 

\begin{equation*}
\label{eq:npt}
\textrm{TPR} = \frac{\sum{\textrm{True Positive}}}{\sum{\textrm{True Positive}}+\sum{\textrm{False Negative}}},
\end{equation*}

and 

\begin{equation*}
\label{eq:npt}
\textrm{FPR} = \frac{\sum{\textrm{False Positive}}}{\sum{\textrm{False Positive}}+\sum{\textrm{True Negative}}}.
\end{equation*}

\noindent
The ROC is generated by evaluating the TPR and FPR at all thresholds of score returned by a classifier.  For example, the ROC is calculated for a probabilistic classifier by varying a threshold from $0.0$ to $1.0$ and calculating FPR and TPR for each value in the range.  Area under the curve (AUC) is a common single metric to compare ROC curves, and as the name implies, is just the area under the ROC curve.  An AUC of 1 is perfect, and an AUC of 0.5 is the same as chance in a binary classifier.

Averaging results over classes is done using both a micro and macro average.  Micro averaging takes into account the number of elements in the test set.  This means that smaller classes will account for less in the average than larger classes.  Macro, on the other hand, averages over all classes regardless of the number of elements in each individual class.  For this paper, macro averaging is probably a better predictor of performance as the distributions of classes in our dataset may not accurately represent the true distributions in the wild.  However, both measures are provided for completeness.

\subsection{Experimental Designs}
The proposed technique is evaluated using three different experimental designs:

\begin{enumerate}
  \item binary classification with random holdout test sets to measure the general ability to detect DGA vs. non-DGA,
  \item binary classification with holdout DGA algorithm families to measure the ability to detect new DGAs, and
  \item multiclass classification to measure the ability to distinguish one DGA algorithm from another. 
\end{enumerate}

The binary classification experimental design tests each DGA classifier for it's ability to make an accurate binary decision: \textit{DGA} or \textit{not DGA}.  The DGA class consists of domains from all thirty families in our training set.  This experiment is run using $n$-fold cross validation with ten folds.  Evaluation is accomplished with both an ROC as well as a detailed Precision, Recall and $F_1$ score broken down by each class.  Both the micro and macro averages of Precision, Recall and $F_1$ score are also given.

In the second experiment, we test each classifier's ability to discover new DGA families not used in the training set.  The ten smallest DGA families are removed from the dataset and each classifier is trained on all samples from the remaining classes.  Precision, Recall and $F_1$ score is calculated on the test set.  In addition, we find both the micro and macro average of these scores over all classes for each algorithm.  

The multiclass classification design tests each DGA classifier for its ability to make an accurate decision on the family of DGA.  The random forest DGA classifier (using manual features) uses a One vs. Rest while the LSTM and Bigram classifiers do a direct multiclass classification.  We display a class breakdown of Precision, Recall and 
$F_1$ score for each class as well as the micro and macro average.  

\subsection{Data}
This paper uses open datasets for reproducibility.  A real-world system should use an expanded dataset to make it more difficult for an adversary to reverse engineer and defeat the classifier.  The experimental designs use data from two sources.

\begin{enumerate}
  \item The \texttt{Alexa} top 1 million domains \cite{Alexa1M} are used for training domains that are not DGAs.
  \item The OSINT DGA feed from Bambenek Consulting \cite{dgafeed} is used for DGA domains.
\end{enumerate}

The OSINT DGA feed consists of thirty families of DGAs with a varying number of examples from each class.  This feed contains approximately 750,000 DGA examples.

\subsection{Comparison to state of the art}
For each experiment, we compare the featureless LSTM DGA classifier to
\begin{itemize}
\item a featureless HMM model\footnote{HMM is excluded from the multiclass experiment due to poor performance.} defined in \cite{antonakakis2012throw},
\item logistic regression on character bigrams (simple features), and
\item a random forest DGA classifier using manually-crafted domain features defined in \cite{antonakakis2012throw, yadav2010detecting, yadav2012detecting, schiavoni2014phoenix}.
\end{itemize}

In particular, the manually crafted features of the random forest DGA classifier include the following:
\begin{itemize}
\item length of domain name,
\item entropy of character distribution in domain name,
\item vowel to consonant ratio,
\item Alexa 1M $n$-gram frequency distribution co-occurrence count, where $n=3,4$ or $5$,
\item \textit{$n$-gram normality score}, and
\item \textit{meaningful characters ratio}.
\end{itemize}
Note that for the \textit{$n$-gram normality score}, we use $n=3$, $n=4$ and $n=5$ as three distinct features as opposed to $n=1$, $n=2$ and $n=3$ as in  \cite{schiavoni2014phoenix} since the larger $n$-gram size performed better in preliminary experiments.  In addition, features were trained in a random forest DGA classifier as opposed to a Mahalanobis distance classifier as used in \cite{schiavoni2014phoenix} as the random forest DGA classifier produced better results.

Four separate HMMs are trained with one trained on the non-DGA class, and three trained on the three largest DGA classes in terms of support (\texttt{Post}, \texttt{banjori}, and \texttt{ramnit}).  The number of hidden states is set to the average length of the domain names in the training set.  We use the Neyman-Pearson likelihood ratio test to classify a domain as DGA generated if 

\begin{equation*}
\label{eq:npt}
\log P_{i^*} - \log P_0 \ge \eta,
\end{equation*}

\noindent
where

\begin{equation*}
\label{eq:npt_argmax}
i^* = \underset{i \in \{\textrm{banjori, ramnit, Post}\}}{\arg\max} P_{i},
\end{equation*}
 
\noindent
$P_0$ is the probability of being a non-DGA, and $\eta$ is a user specified threshold.  There are a few key differences from the HMM presented in \cite{antonakakis2012throw}.  Authors in \cite{antonakakis2012throw} use a distinct HMM for each family of DGA, while we only create an HMM for the three largest classes of DGAs in the training set.  In addition, we use the Neyman-Pearson likelihood ratio test as opposed to a threshold directly on the maximum HMM score from the DGA HMMs. Preliminary results showed a significant improvement in ROC over the algorithm presented in \cite{antonakakis2012throw} when using these updates.

Even with the improved algorithm, the HMM performed worse than other techniques evaluated in this paper.  This is especially true for the multiclass experiment.  The original HMM algorithm in \cite{antonakakis2012throw} was presented on only four classes, each with a significant support.  This is unlike our setup that has thirty classes with varying degrees of support.  For this reason we omit HMM results for the multiclass experiment.
    
We also compare our results with those of retrospective techniques as reported in \cite{krishnan2013crossing}.  This comparison is only done for the binary classification as our dataset only contains domain names without any contextual information. In addition, retrospective techniques perform far worse than real-time techniques for binary classification and, therefore, will likely degrade even further for multiclass classification.

\section{Results}
\label{results}
Results for the three experiments and an interpretation of model performance are presented in this section.
\subsection{Binary Classification}
\begin{figure}
\centering
\includegraphics[scale=0.25]{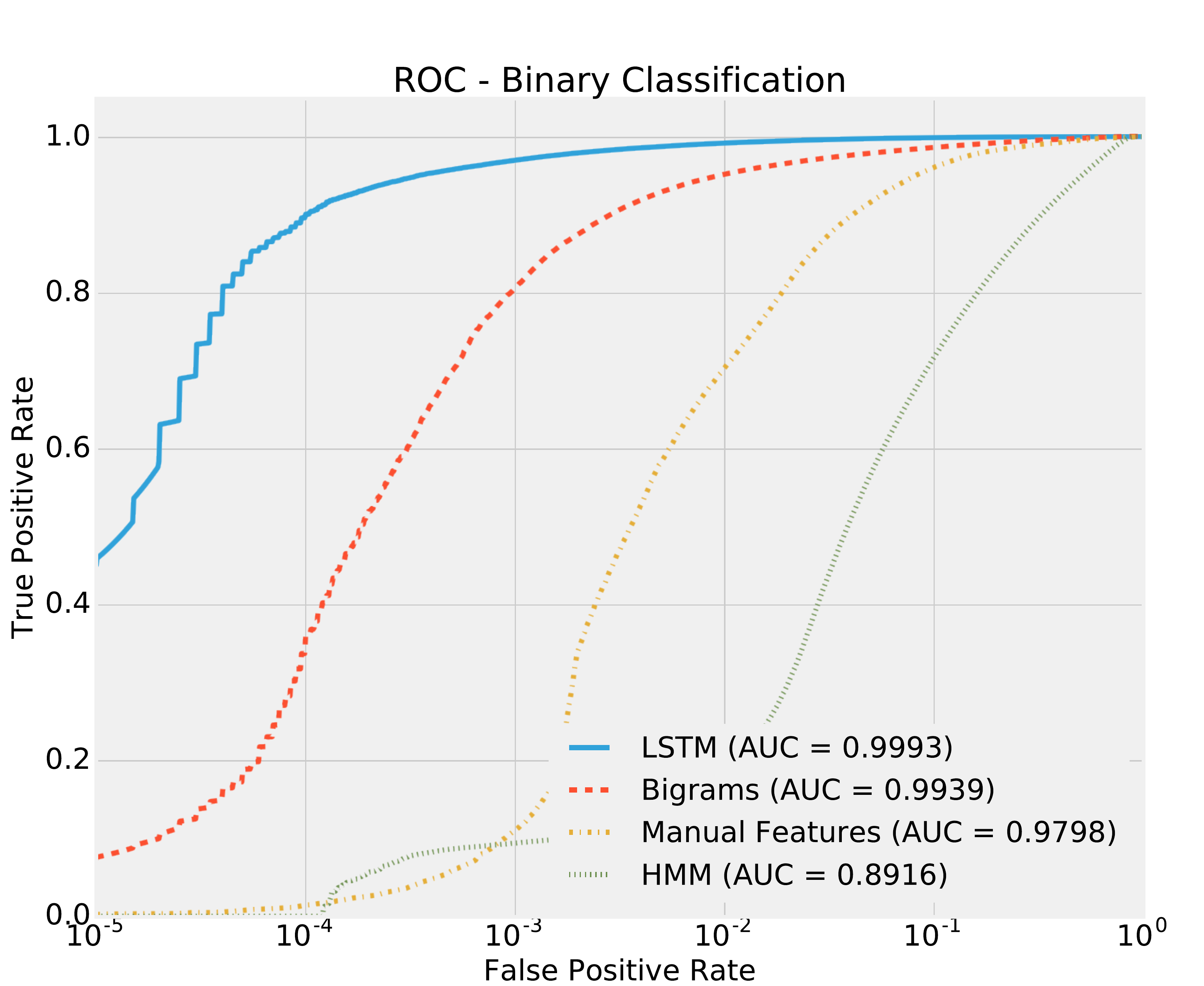}
\caption{ROC curves for binary classification of DGA and non-DGA generated domains using the LSTM model, logistic regression with bigram features, random forest classifier with manual features, and HMM classifier.}
\label{fig:binary_roc}
\end{figure}

\begin{table}
\caption{True Positive Rates of LSTM compared to Retrospective techniques}
\label{table:retrospective_tbl}
\begin{center}
\resizebox{\columnwidth}{!}{%
\begin{tabular}{l||c|c|}
\textbf{Technique} & \textbf{True Positive Rate} & \textbf{False Positive Rate} \T \\
\hline
\rowcolor{Gray}KL Divergence \cite{yadav2010detecting, yadav2012detecting} & $<0.5$ & 0.05 \T \\
NXDomains \cite{krishnan2013crossing} & 0.94 & 0.002 \T \\
\rowcolor{Gray}LSTM & \textbf{0.98} & \textbf{0.001} \B \\
\end{tabular}
}
\end{center}
\end{table}

\begin{table*}[ht]
\caption{Precision, Recall and $F_1$ Score for Binary Classifiers}
\label{table:binary_prf}
\begin{center}
\resizebox{\textwidth}{!}{%
\begin{tabular}{l||c|c|c|c||c|c|c|c||c|c|c|c||c|}
\multicolumn{1}{l||}{\textbf{Domain Type}} & \multicolumn{4}{c||}{\textbf{Precision}}  & \multicolumn{4}{c||}{\textbf{Recall}}  & \multicolumn{4}{c||}{\textbf{$\boldsymbol{F_1}$ Score}} & \multicolumn{1}{c|}{\textbf{Support}} \T\B \\
 & HMM & Features & Bigram & LSTM & HMM & Features & Bigram & LSTM & HMM & Features & Bigram & LSTM & \T \\
\hline
\texttt{Alexa} & 0.8300 & 0.9400 & 0.9700 & 0.9900 & 1.0000 & 1.0000 & 1.0000 & 1.0000 & 0.9100 & 0.9700 & 0.9900 & 0.9900 & 300064 \T \\
\rowcolor{Gray}\texttt{Cryptolocker} & 1.0000 & 1.0000 & 1.0000 & 1.0000 & 0.9000 & 0.9800 & 0.9700 & 0.9900 & 0.9500 & 0.9900 & 0.9900 & 0.9900 & 1799 \\
\texttt{P2P Gameover Zeus} & 1.0000 & 1.0000 & 1.0000 & 1.0000 & 0.9900 & 1.0000 & 1.0000 & 1.0000 & 0.9900 & 1.0000 & 1.0000 & 1.0000 & 298 \\
\rowcolor{Gray}\texttt{Post Tovar GOZ} & 1.0000 & 1.0000 & 1.0000 & 1.0000 & 1.0000 & 1.0000 & 1.0000 & 1.0000 & 1.0000 & 1.0000 & 1.0000 & 1.0000 & 19863 \\
\texttt{Volatile Cedar / Explosive} & 0.0000 & 1.0000 & 1.0000 & 1.0000 & 0.0000 & 0.4600 & 0.4900 & 0.9900 & 0.0000 & 0.6300 & 0.6600 & 1.0000 & 294 \\
\rowcolor{Gray}\texttt{banjori} & 1.0000 & 1.0000 & 1.0000 & 1.0000 & 0.5900 & 0.9400 & 1.0000 & 1.0000 & 0.7400 & 0.9700 & 1.0000 & 1.0000 & 121678 \\
\texttt{bedep} & 1.0000 & 1.0000 & 1.0000 & 1.0000 & 0.8100 & 1.0000 & 1.0000 & 1.0000 & 0.8900 & 1.0000 & 1.0000 & 1.0000 & 53 \\
\rowcolor{Gray}\texttt{beebone} & 0.0000 & 1.0000 & 1.0000 & 1.0000 & 0.0000 & 1.0000 & 0.9700 & 1.0000 & 0.0000 & 1.0000 & 0.9900 & 1.0000 & 65 \\
\texttt{corebot} & 1.0000 & 1.0000 & 1.0000 & 1.0000 & 0.5900 & 1.0000 & 1.0000 & 0.9600 & 0.7400 & 1.0000 & 1.0000 & 0.9800 & 81 \\
\rowcolor{Gray}\texttt{cryptowall} & 1.0000 & 1.0000 & 1.0000 & 1.0000 & 0.1100 & 0.0600 & 0.1400 & 0.1200 & 0.1900 & 0.1100 & 0.2500 & 0.2100 & 29 \\
\texttt{dircrypt} & 1.0000 & 1.0000 & 1.0000 & 1.0000 & 0.9100 & 0.9200 & 0.9600 & 0.9600 & 0.9500 & 0.9600 & 0.9800 & 0.9800 & 150 \\
\rowcolor{Gray}\texttt{dyre} & 1.0000 & 1.0000 & 1.0000 & 1.0000 & 1.0000 & 1.0000 & 0.9900 & 1.0000 & 1.0000 & 1.0000 & 0.9900 & 1.0000 & 2389 \\
\texttt{fobber} & 1.0000 & 1.0000 & 1.0000 & 1.0000 & 0.8900 & 0.9600 & 0.9700 & 0.9700 & 0.9400 & 0.9800 & 0.9800 & 0.9900 & 181 \\
\rowcolor{Gray}\texttt{geodo} & 1.0000 & 1.0000 & 1.0000 & 1.0000 & 0.9100 & 1.0000 & 0.9900 & 0.9900 & 0.9500 & 1.0000 & 1.0000 & 1.0000 & 173 \\
\texttt{hesperbot} & 1.0000 & 1.0000 & 1.0000 & 1.0000 & 0.8300 & 0.7700 & 0.8500 & 0.9700 & 0.9100 & 0.8700 & 0.9200 & 0.9800 & 58 \\
\rowcolor{Gray}\texttt{matsnu} & 0.0000 & 0.0000 & 0.0000 & 0.0000 & 0.0000 & 0.0000 & 0.0000 & 0.0000 & 0.0000 & 0.0000 & 0.0000 & 0.0000 & 14 \\
\texttt{murofet} & 1.0000 & 1.0000 & 1.0000 & 1.0000 & 0.9200 & 1.0000 & 0.9900 & 1.0000 & 0.9600 & 1.0000 & 1.0000 & 1.0000 & 4292 \\
\rowcolor{Gray}\texttt{necurs} & 1.0000 & 1.0000 & 1.0000 & 1.0000 & 0.8800 & 0.8400 & 0.9400 & 0.9600 & 0.9400 & 0.9100 & 0.9700 & 0.9800 & 1232 \\
\texttt{nymaim} & 1.0000 & 1.0000 & 1.0000 & 1.0000 & 0.8000 & 0.5600 & 0.7300 & 0.8000 & 0.8900 & 0.7200 & 0.8500 & 0.8900 & 1815 \\
\rowcolor{Gray}\texttt{pushdo} & 1.0000 & 1.0000 & 1.0000 & 1.0000 & 0.6600 & 0.4700 & 0.5600 & 0.6000 & 0.7900 & 0.6400 & 0.7200 & 0.7500 & 507 \\
\texttt{pykspa} & 1.0000 & 1.0000 & 1.0000 & 1.0000 & 0.7200 & 0.5400 & 0.7700 & 0.9000 & 0.8400 & 0.7000 & 0.8700 & 0.9500 & 4250 \\
\rowcolor{Gray}\texttt{qakbot} & 1.0000 & 1.0000 & 1.0000 & 1.0000 & 0.9100 & 0.9600 & 0.9600 & 0.9800 & 0.9500 & 0.9800 & 0.9800 & 0.9900 & 1517 \\
\texttt{ramnit} & 1.0000 & 1.0000 & 1.0000 & 1.0000 & 0.8800 & 0.9100 & 0.9400 & 0.9600 & 0.9400 & 0.9500 & 0.9700 & 0.9800 & 27439 \\
\rowcolor{Gray}\texttt{ranbyus} & 1.0000 & 1.0000 & 1.0000 & 1.0000 & 0.9000 & 1.0000 & 0.9800 & 0.9800 & 0.9500 & 1.0000 & 0.9900 & 0.9900 & 2625 \\
\texttt{shifu} & 1.0000 & 1.0000 & 1.0000 & 1.0000 & 0.7200 & 0.2100 & 0.6600 & 0.7700 & 0.8400 & 0.3500 & 0.8000 & 0.8700 & 697 \\
\rowcolor{Gray}\texttt{shiotob/urlzone/bebloh} & 1.0000 & 1.0000 & 1.0000 & 1.0000 & 0.9000 & 0.9700 & 0.9500 & 0.9800 & 0.9500 & 0.9900 & 0.9700 & 0.9900 & 3031 \\
\texttt{simda} & 1.0000 & 1.0000 & 1.0000 & 1.0000 & 0.5600 & 0.0800 & 0.4000 & 0.9200 & 0.7100 & 0.1400 & 0.5800 & 0.9600 & 4449 \\
\rowcolor{Gray}\texttt{suppobox} & 1.0000 & 0.0000 & 1.0000 & 1.0000 & 0.0100 & 0.0000 & 0.0000 & 0.3200 & 0.0200 & 0.0000 & 0.0100 & 0.4800 & 298 \\
\texttt{symmi} & 0.0000 & 1.0000 & 1.0000 & 1.0000 & 0.0000 & 1.0000 & 0.7900 & 0.6900 & 0.0000 & 1.0000 & 0.8800 & 0.8200 & 18 \\
\rowcolor{Gray}\texttt{tempedreve} & 1.0000 & 1.0000 & 1.0000 & 1.0000 & 0.7600 & 0.5700 & 0.8500 & 0.7700 & 0.8600 & 0.7300 & 0.9200 & 0.8700 & 74 \\
\texttt{tinba} & 1.0000 & 1.0000 & 1.0000 & 1.0000 & 0.8900 & 0.9800 & 0.9700 & 0.9900 & 0.9400 & 0.9900 & 0.9900 & 0.9900 & 18505 \B \\
\hline
Micro Average & 0.9008 & 0.9647 & 0.9826 & \textbf{0.9942} & 0.8815 & 0.9639 & 0.9848 & \textbf{0.9937} & 0.8739 & 0.9593 & 0.9851 & \textbf{0.9906} & 16708 \T \\
Macro Average & 0.8655 & 0.9335 & 0.9668 & \textbf{0.9674} & 0.6787 & 0.7477 & 0.8006 & \textbf{0.8571} & 0.7335 & 0.7929 & 0.8468 & \textbf{0.8913} & 16708 \B \\
\end{tabular}
}
\end{center}
\end{table*}

The ROC curves for the HMM, random forest classifier with manually-crafted features (Manual Features), logistic regression classifier on character bigrams (Bigrams), and LSTM DGA clasifier (LSTM) are presented in Fig. \ref{fig:binary_roc}.  Note that the abscissa (false positive rate) is on a log scale to highlight the differences in the algorithms.  LSTM provides the best performance with an AUC of 0.9993 with the bigram model at 0.9939.  The difference between the two algorithms may seem small, but are actually quite significant in a production system.  As an example, the LSTM model can classify 90\% of all DGAs with a 1 in 10,000 false positive rate.  On the other hand, a Bigram model will classify the same percentage of DGA's with a 1 in 550 false positive rate (i.e., the Bigram model produces a false positive rate that is 20$\times$ that of the LSTM model).

The breakdown of Precision, Recall, and $F_1$ for each class as classified by the binary classifiers is given in Table \ref{table:binary_prf}.  The support (size of test set) is given in the last column.  In general, classes that are the most difficult to detect have smaller support.  This is expected as they have a smaller contribution to model updates during training than larger classes.  In addition \texttt{matsnu} was undetectable by all algorithms.  \texttt{matsnu} is a dictionary-based DGA, meaning it is created by randomly selecting and concatenating multiple words from a dictionary.  Interestingly, \texttt{suppobox} is also a dictionary based DGA, but was detectable (to some extent) by the LSTM.  The size of the \texttt{suppobox} training was about twenty times that of \texttt{matsnu} allowing for repeats of randomly selected dictionary words.  These repeats allow the LSTM to learn the dictionaries of such DGAs.  We leave an in-depth analysis of dictionary based DGA to future work.

The HMM performed worse than expected.  The results presented in \cite{antonakakis2012throw} only used a small number of homogenous DGA families (\texttt{Conficker}, \texttt{Murofet}, \texttt{Bobax} and, \texttt{Sinowal}) while the experiments in this paper use over 30 different families.  Some of these families in this paper are related, but overall, our results were generated from a larger/more rich dataset.  As discussed later in this paper, the letter distributions are very different across the 30 DGA families used in this paper. For example, DGA families such as \texttt{Cryptolocker} and \texttt{ramnit} have near uniform distributions over letters, \texttt{dyre} has a uniform distribution over hexadecimal characters with a dictionary word as a prefix, and \texttt{suppobox} and \texttt{matsnu} use English words to create domains giving a distribution very similar to english based domains.  In contrast, \texttt{Conficker} \cite{porras2009foray}, \texttt{Murofet} \cite{andriesse2013highly}, \texttt{Bobax} \cite{kraken} and \texttt{Sinowal} \cite{stone2009your} all use a generator that gives a uniform distribution over letters similar to \texttt{Cryptolocker} and \texttt{ramnit}.

Table \ref{table:retrospective_tbl} displays the true positive rate and false positive rate for retrospective techniques as compared to the LSTM technique presented by this paper.  As can be seen, the LSTM technique significantly outperforms the best retrospective techniques.



\subsection{Leave-Class-Out Binary Classification}
\begin{table}
\caption{Recall for all leave-out classes}
\label{table:binary_loo}
\begin{center}
\resizebox{\columnwidth}{!}{%
\begin{tabular}{l||c|c|c|c||c}
\textbf{Domain Type} & \textbf{HMM} & \textbf{Features} & \textbf{Bigram} & \textbf{LSTM} & \textbf{Support} \T \\
\hline
\texttt{bedep} & 0.83 & \textbf{0.99} & \textbf{0.99} & \textbf{0.99} & 172 \T \\
\rowcolor{Gray}\texttt{beebone} & 0.00 & \textbf{1.00} & 0.00 & 0.00 & 210 \T \\
\texttt{corebot} & 0.59 & \textbf{1.00} & 0.71 & 0.77 & 280 \T \\
\rowcolor{Gray}\texttt{cryptowall} & \textbf{0.30} & 0.20 & 0.18 & 0.20 & 94 \T \\
\texttt{dircrypt} & 0.94 & 0.91 & 0.94 & \textbf{0.97} & 510 \T \\
\rowcolor{Gray}\texttt{fobber} & 0.93 & 0.93 & 0.95 & \textbf{0.99} & 600 \T \\
\texttt{hesperbot} & 0.90 & 0.76 & 0.86 & \textbf{0.92} & 192 \T \\
\rowcolor{Gray}\texttt{matsnu} & 0.00 & 0.02 & \textbf{0.04} & 0.0 & 48 \T \\
\texttt{symmi} & 0.00 & \textbf{1.00} & 0.11 & 0.06 & 64 \T \\
\rowcolor{Gray}\texttt{tempedreve} & 0.81 & 0.61 & 0.80 & \textbf{0.84} & 249 \B \\
\hline
micro & 0.78 & \textbf{0.90} & 0.80 & 0.81 & \multicolumn{1}{c} \T \\
macro & 0.53 & \textbf{0.74} & 0.558 & 0.642 & \multicolumn{1}{c} \B \\
\end{tabular}
}
\end{center}
\end{table}

The binary leave-one-out classifier is interesting as it tests each algorithm's robustness to DGA families not seen during training.  Only Recall is presented for this experiment as there are no non-DGA generated domains in this test set.  The results for this experiment are shown in Table \ref{table:binary_loo}. 

\sloppy
The manual features random forest classifier performs best in terms of both micro and macro average.  On the other hand, the LSTM classifier has the most families that it performs best on (five in total as opposed to four in total for the manual features classifier).  The biggest discrepancy between manual features and LSTM was with \texttt{beebone}.  In particular, the manual features classifier identifies all of the \texttt{beebone} samples, while the LSTM model recovers none.  The domain names from \texttt{beebone} have a rigid structure, like \texttt{ns1.backdates13.biz} and\texttt{ns1.backdates0.biz}, so that the LSTM model was unable to learn the structure that included the word \texttt{backdates} without training data.  The results are nearly as dramatic for \texttt{symmi}, which produces nearly-pronounceable domain names like \texttt{hakueshoubar.ddns.net}, by drawing a random vowel or a random consonant at each even-numbered index, then drawing a random character of the opposite class (vowel/consonant) in the subsequent index location.  These examples highlight blind spots in the LSTM classifier.  However, these blind spots can be easily fixed through training with the use of an adversarial network (i.e., train a generator network that creates domains that confuses our classifier).

Apparently, the structure of some DGA families--even if not elaborately designed--are peculiar enough to necessitate their inclusion in the training set.  As evident in the results for Experiment 1 in Table \ref{table:binary_prf}, the LSTM readily detects these families with distinct structure when accounted for in the training set with sufficient support.  The manual features appear to be generic enough to detect these families with high recall.  However, its important to note that manual features were designed specifically for known DGA families and all of our DGAs in our test set are known (i.e., our dataset is known and labeled) making this experiment biased to a feature based classifier.  Even with this bias, the LSTM classifier still performs best in terms of the number of DGA families it detects.


\subsection{Multiclass}
\begin{table*}[ht]
\caption{Precision, Recall and $F_1$ Score for Multiclass Classifiers}
\label{table:multiclass_prf}
\begin{center}
\resizebox{\textwidth}{!}{%
\begin{tabular}{l||c|c|c||c|c|c||c|c|c||c}
\multicolumn{1}{c||}{} & \multicolumn{3}{c||}{\textbf{Precision}}  & \multicolumn{3}{c||}{\textbf{Recall}}  & \multicolumn{3}{c||}{\textbf{$\boldsymbol{F_1}$ Score}} & \multicolumn{1}{c}{} \T\B \\
Domain Type & Features & Bigram &  LSTM & Features & Bigram &  LSTM & Features & Bigram &  LSTM & Support \T \\
\hline
\texttt{Alexa} & 0.914 & 0.980 & \textbf{0.990} & 0.960 & 0.990 & \textbf{1.000} & 0.940 & 0.988 & \textbf{0.990} & 199978 \T \\
\rowcolor{Gray}\texttt{Cryptolocker} & 0.000 & 0.000 & 0.000 & 0.000 & 0.000 & 0.000 & 0.000 & 0.000 & 0.000 & 1189 \\
\texttt{P2P Gameover Zeus} & 0.000 & \textbf{0.343} & 0.327 & 0.000 & \textbf{0.288} & 0.217 & 0.000 & \textbf{0.308} & 0.247 & 196 \\
\rowcolor{Gray}\texttt{Post Tovar GOZ} & 0.941 & \textbf{1.000} & \textbf{1.000} & \textbf{1.000} & \textbf{1.000} & \textbf{1.000} & 0.970 & \textbf{1.000} & \textbf{1.000} & 13185 \\
\texttt{Volatile Cedar / Explosive} & 0.000 & \textbf{1.000} & 0.987 & 0.000 & \textbf{1.000} & 0.980 & 0.000 & \textbf{1.000} & 0.980 & 200 \\
\rowcolor{Gray}\texttt{banjori} & 0.900 & 0.990 & \textbf{1.000} & 0.938 & \textbf{1.000} & \textbf{1.000} & 0.920 & \textbf{1.000} & \textbf{1.000} & 81281 \\
\texttt{bedep} & 0.000 & 0.000 & \textbf{0.943} & 0.000 & 0.000 & \textbf{0.107} & 0.000 & 0.000 & \textbf{0.187} & 34 \\
\rowcolor{Gray}\texttt{beebone} & \textbf{1.000} & \textbf{1.000} & \textbf{1.000} & 0.560 & \textbf{1.000} & \textbf{1.000} & 0.713 & \textbf{1.000} & \textbf{1.000} & 42 \\
\texttt{corebot} & 0.000 & \textbf{1.000} & \textbf{1.000} & 0.000 & 0.980 & \textbf{0.990} & 0.000 & 0.990 & \textbf{0.993} & 54 \\
\rowcolor{Gray}\texttt{cryptowall} & 0.000 & 0.000 & 0.000 & 0.000 & 0.000 & 0.000 & 0.000 & 0.000 & 0.000 & 15 \\
\texttt{dircrypt} & 0.000 & \textbf{0.083} & 0.000 & 0.000 & \textbf{0.010} & 0.000 & 0.000 & \textbf{0.020} & 0.000 & 100 \\
\rowcolor{Gray}\texttt{dyre} & 0.985 & 0.988 & \textbf{1.000} & \textbf{1.000} & 0.988 & \textbf{1.000} & 0.991 & 0.988 & \textbf{1.000} & 1600 \\
\texttt{fobber} & 0.000 & 0.000 & \textbf{0.177} & 0.000 & 0.000 & \textbf{0.023} & 0.000 & 0.000 & \textbf{0.040} & 121 \\
\rowcolor{Gray}\texttt{geodo} & 0.000 & 0.000 & 0.000 & 0.000 & 0.000 & 0.000 & 0.000 & 0.000 & 0.000 & 114 \\
\texttt{hesperbot} & 0.000 & 0.000 & 0.000 & 0.000 & 0.000 & 0.000 & 0.000 & 0.000 & 0.000 & 36 \\
\rowcolor{Gray}\texttt{matsnu} & 0.000 & 0.000 & 0.000 & 0.000 & 0.000 & 0.000 & 0.000 & 0.000 & 0.000 & 9 \\
\texttt{murofet} & \textbf{0.883} & 0.643 & 0.783 & 0.066 & 0.542 & \textbf{0.700} & 0.122 & 0.590 & \textbf{0.737} & 2845 \\
\rowcolor{Gray}\texttt{necurs} & 0.000 & 0.000 & \textbf{0.643} & 0.000 & 0.000 & \textbf{0.093} & 0.000 & 0.000 & \textbf{0.160} & 827 \\
\texttt{nymaim} & 0.000 & 0.390 & \textbf{0.477} & 0.000 & 0.113 & \textbf{0.190} & 0.000 & 0.175 & \textbf{0.267} & 1222 \\
\rowcolor{Gray}\texttt{pushdo} & 0.000 & 0.770 & \textbf{0.853} & 0.000 & 0.588 & \textbf{0.640} & 0.000 & 0.665 & \textbf{0.730} & 339 \\
\texttt{pykspa} & 0.000 & 0.788 & \textbf{0.910} & 0.000 & 0.593 & \textbf{0.713} & 0.000 & 0.675 & \textbf{0.800} & 2827 \\
\rowcolor{Gray}\texttt{qakbot} & 0.000 & \textbf{0.590} & 0.590 & 0.000 & 0.232 & \textbf{0.387} & 0.000 & 0.338 & \textbf{0.463} & 993 \\
\texttt{ramnit} & 0.566 & 0.637 & \textbf{0.770} & 0.654 & 0.763 & \textbf{0.850} & 0.605 & 0.690 & \textbf{0.810} & 18308 \\
\rowcolor{Gray}\texttt{ranbyus} & 0.439 & 0.000 & \textbf{0.450} & 0.000 & 0.000 & \textbf{0.517} & 0.001 & 0.000 & \textbf{0.460} & 1736 \\
\texttt{shifu} & 0.000 & 0.037 & \textbf{0.560} & 0.000 & 0.003 & \textbf{0.570} & 0.000 & 0.007 & \textbf{0.553} & 465 \\
\rowcolor{Gray}\texttt{shiotob/urlzone/bebloh} & 0.000 & 0.965 & \textbf{0.973} & 0.000 & 0.853 & \textbf{0.907} & 0.000 & 0.907 & \textbf{0.940} & 2016 \\
\texttt{simda} & 0.000 & 0.840 & \textbf{0.930} & 0.000 & 0.750 & \textbf{0.977} & 0.000 & 0.792 & \textbf{0.950} & 2955 \\
\rowcolor{Gray}\texttt{suppobox} & 0.000 & 0.392 & \textbf{0.833} & 0.000 & 0.062 & \textbf{0.517} & 0.000 & 0.112 & \textbf{0.627} & 197 \\
\texttt{symmi} & 0.000 & 0.625 & \textbf{0.913} & 0.000 & 0.117 & \textbf{0.857} & 0.000 & 0.200 & \textbf{0.883} & 11 \\
\rowcolor{Gray}\texttt{tempedreve} & 0.000 & \textbf{0.043} & 0.000 & 0.000 & \textbf{0.010} & 0.000 & 0.000 & \textbf{0.018} & 0.000 & 50  \\
\texttt{tinba} & 0.821 & 0.735 & \textbf{0.910} & 0.923 & 0.802 & \textbf{0.990} & 0.869 & 0.767 & \textbf{0.950} & 12332 \B \\
\hline
Micro Average & 0.851 & 0.933 & \textbf{0.963} & 0.888 & 0.944 & \textbf{0.970} & 0.867 & 0.940 & \textbf{0.963} & 11138 \T \\
Macro Average & 0.240 & 0.479 & \textbf{0.614} & 0.197 & 0.409 & \textbf{0.523} & 0.198 & 0.427 & \textbf{0.541} & 11138 \B \\
\end{tabular}
}
\end{center}
\end{table*}

\begin{figure}
\includegraphics[scale=0.45]{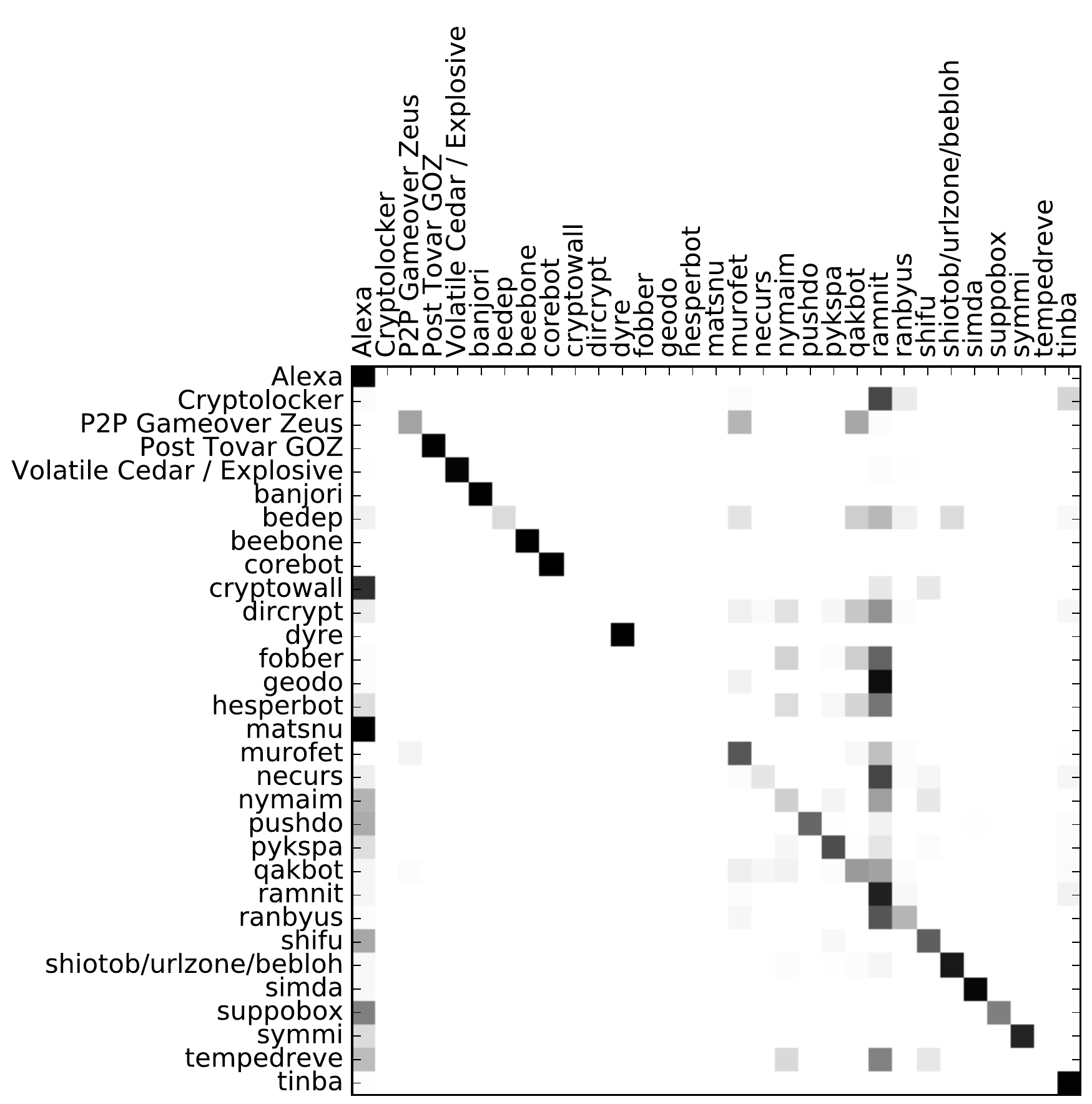}
\caption{Confusion matrix for the LSTM multiclass model.  Blocks represent the fraction of DGA families on the vertical axis classified as DGA families on the horizontal axis, where 0 is depicted as white and 1 depicted as black.  A perfect classifier would produce an identity matrix composed of black blocks.}
\label{table:confusion}
\end{figure}

\begin{figure}
\includegraphics[scale=0.41]{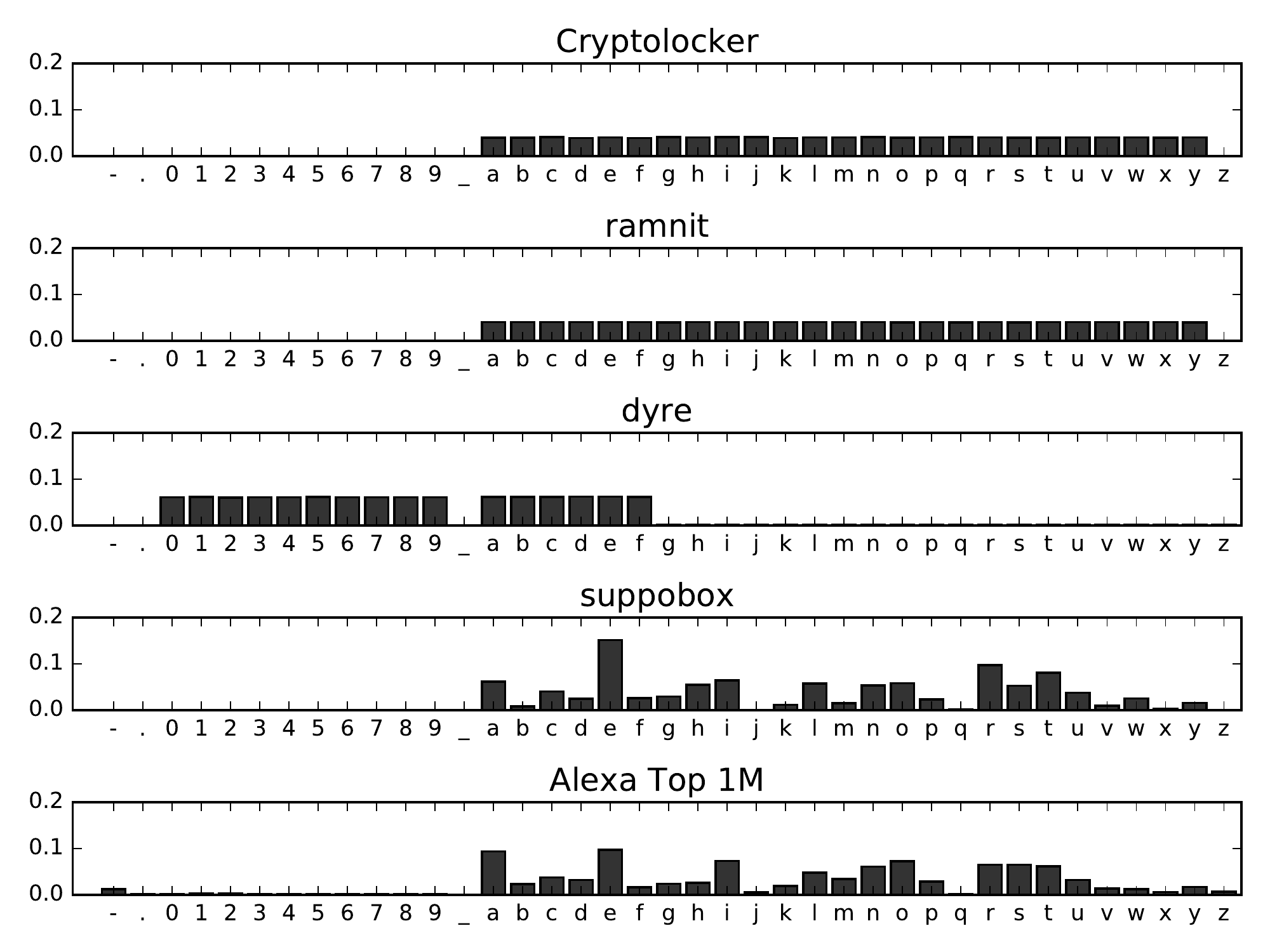}
\caption{Unigram distributions for \texttt{Cryptolocker}, \texttt{ramnit}, \texttt{dyre}, \texttt{suppobox} and the \texttt{Alexa} top one million.}
\label{table:unigram_distro}
\end{figure}

The HMM results were omitted from the multiclass experiments due to poor performance.  As stated previously, the HMM algorithm was designed for few DGAs, whereas our experiments include over 30 classes.  Precision, Recall, and $F_1$ is displayed in Table \ref{table:multiclass_prf} for the random forest classifier with manual features (Manual Features), multinomial logistic regression on character bigrams (Bigram) and the LSTM classifier.    The LSTM classifier significantly outperforms the other two algorithms in both the micro and macro averaged Precision, Recall, and $F_1$ score.  In general, poor performance resulted from classes with small representation.  One exception was \texttt{Cryptolocker}, which no multiclass classifier was able to detect.  However, all the binary classifiers were able to distinguish \texttt{Cryptolocker} from other families.  

Fig. \ref{table:confusion} shows the confusion matrix for the LSTM multiclass classifier.  A large number of the incorrectly classified \texttt{Cryptolocker} DGAs are classified as \texttt{ramnit}.  To further investigate, the unigram distributions for four DGA families and Alexa are shown in Fig. \ref{table:unigram_distro}.  The distributions for \texttt{Cryptolocker} and \texttt{ramnit} are both uniform over the same range.  This is expected as they are both generated using a series of multiplies, divisions and modulos based on a single seed \cite{w32ramnit, cryptolocker}.  On the other hand, \texttt{suppobox} is interesting as it generates unigrams similar to distributions seen by the Alexa top one million domains and is often confused with the benign set.  As discussed earlier, \texttt{suppobox} is an English dictionary-based DGA, meaning domains are constructed by concatenating multiple, randomly chosen words from the English dictionary. Interestingly, only the LSTM classifier was able to consistently detect \texttt{suppobox} (as seen in Table \ref{table:binary_prf}).  This shows LSTM's ability to extract some deep understanding that is lost by other classifiers.  Specifically, the LSTM actually learns the dictionary used by \texttt{suppobox} to construct domains.

\begin{figure}
\includegraphics[scale=0.47]{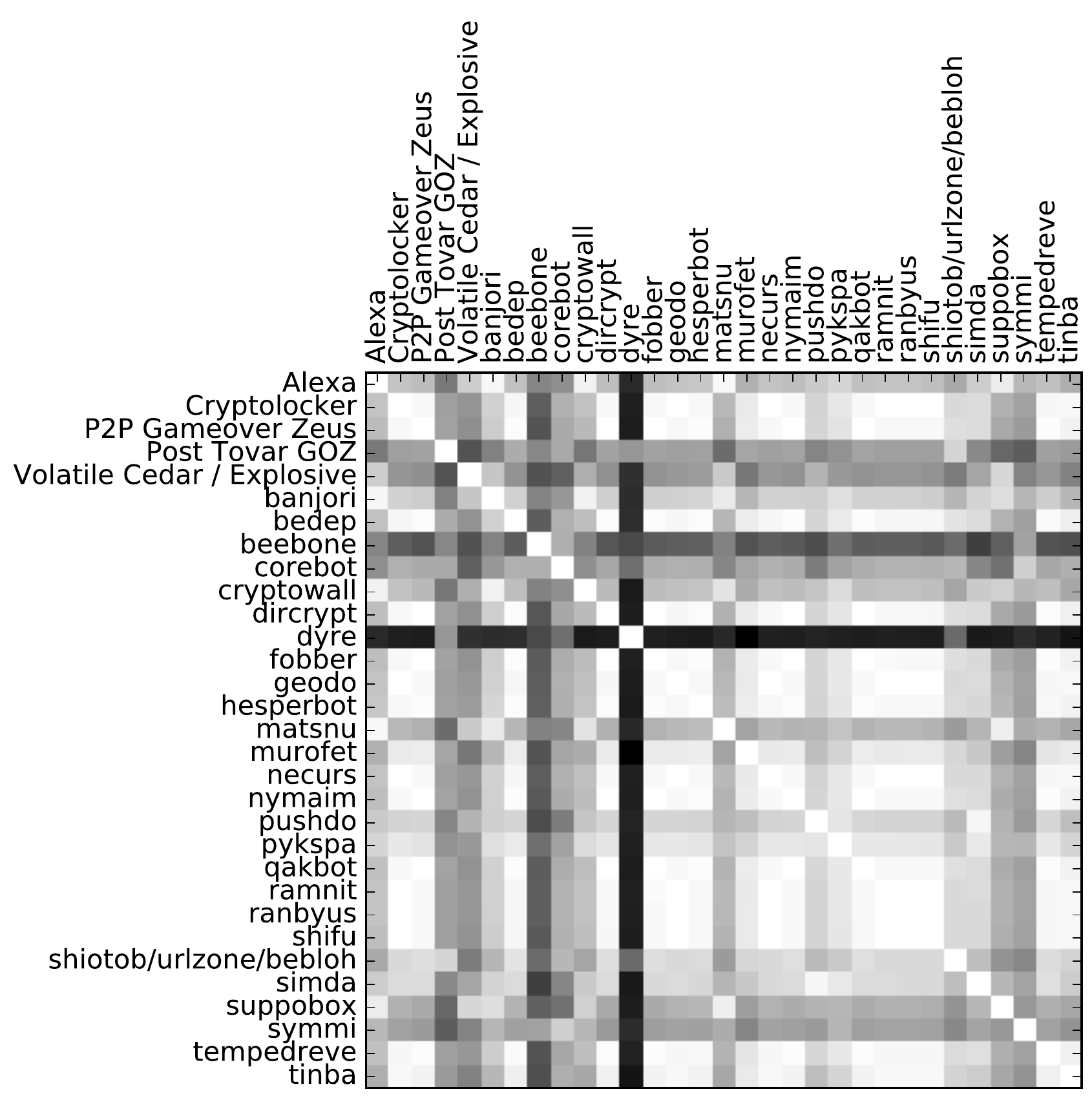}
\caption{All-to-all cosine distance comparison of unigram distributions of all DGA familes and the \texttt{Alexa} top one million.  Distances range from 0 to 1 with 0 depicted as white and 1 depicted as black.}
\label{table:cosine}
\end{figure}

Fig. \ref{table:cosine} shows the all-to-all cosine distance of the unigram distribution between all DGA families and the Alexa top one million domains.  \texttt{dyre} stands out as it is extremely dissimilar to other algorithms.  This is not surprising when comparing this figure to Table \ref{table:unigram_distro}.  \texttt{dyre} has a nearly uniform distribution over primarily hexadecimal numbers (non-hexadecimal letters exist, but are rare).

When comparing both Fig. \ref{table:confusion}, Fig. \ref{table:cosine}, and Table \ref{table:binary_prf}, some correlation can be seen between the unigram distribution and DGA algorithms that are often misclassified.  This suggests that it's not only the lack of representation of these algorithms in the training set, but also the distribution of letters that is causing much of the misclassification.  More specifically, many DGAs produce domains that look nearly identical in terms of their character distributions making multiclass classification difficult if not impossible.  To test this, we performed agglomerative clustering on each DGA's family unigram distribution using cosine distance.  We set a threshold of $0.2$ to define super families (the threshold was chosen using domain knowledge of DGA families).  These super families are shown in Table \ref{table:super_families}.  Interesting super families include Super Family 4 (dictionary-based DGAs), Super Family 5 (randomly selected character DGAs), and Super Family 7 (randomly selected characters with near equal vowels and consonants). 

The same multiclass classification experiment was run on these super families and the results are shown in \ref{table:multiclass_prf_by_family}.  As expected, all three classifiers performed much better on super families.  Results demonstrate that an actual deployment of a multiclass DGA classification would be best run on super families, often alerting on groups of DGAs instead of alerting on a single family. Again, the LSTM classifier performs significantly better than other algorithms.

\begin{table}
\begin{center}
\caption{DGA Super Families}
\label{table:super_families}
\begin{tabular}{ l|p{60mm} } 
 Super Family & Member Families \\
 \hline
 Super Family 0 & dyre \\ 
 \rowcolor{Gray}Super Family 1 & beebone \\ 
 Super Family 2 & Volatile Cedar / Explosive \\ 
 \rowcolor{Gray}Super Family 3 & shiotob/urlzone/bebloh \\ 
 Super Family 4 & banjori, cryptowall, matsnu, suppobox \\ 
 \rowcolor{Gray}Super Family 5 & murofet, tinba, shifu, geodo, necurs, Cryptolocker, ramnit, ranbyus, bedep, hesperbot, tempedreve, fobber, nymaim, qakbot, P2P Gameover Zeus, dircrypt \\ 
 Super Family 6 & pykspa \\ 
 \rowcolor{Gray}Super Family 7 & pushdo, simda \\
 Super Family 8 & Post Tovar GOZ \\ 
 \rowcolor{Gray}Super Family 9 & corebot \\ 
 Super Family 10 & symmi \\  
\end{tabular}
\end{center}
\end{table}


\begin{table*}[ht]
\caption{Precision, Recall and $F_1$ Score for Multiclass Classifiers}
\label{table:multiclass_prf_by_family}
\begin{center}
\resizebox{\textwidth}{!}{%
\begin{tabular}{l||c|c|c||c|c|c||c|c|c||c}
\multicolumn{1}{c||}{} & \multicolumn{3}{c||}{\textbf{Precision}}  & \multicolumn{3}{c||}{\textbf{Recall}}  & \multicolumn{3}{c||}{\textbf{$\boldsymbol{F_1}$ Score}} & \multicolumn{1}{c}{} \T\B \\
Domain Type & Features & Bigram &  LSTM & Features & Bigram &  LSTM & Features & Bigram &  LSTM & Support \T \\
\hline
Alexa & 0.930 & 0.980 & \textbf{0.990} & 0.960 & 0.990 & \textbf{1.000} & 0.940 & \textbf{0.990} & \textbf{0.990} & 199906 \T \\
Super Family 0 & 0.980 & 0.990 & \textbf{1.000} & \textbf{1.000} & 0.990 & \textbf{1.000} & 0.990 & 0.990 & \textbf{1.000} & 1603 \\
Super Family 1 & \textbf{1.000} & \textbf{1.000} & \textbf{1.000} & 0.590 & \textbf{1.000} & \textbf{1.000} & 0.740 & \textbf{1.000} & \textbf{1.000} & 43 \\
Super Family 2 & 0.000 & \textbf{1.000} & \textbf{1.000} & 0.000 & \textbf{1.000} & 0.970 & 0.000 & \textbf{1.000} & 0.990 & 203 \\
Super Family 3 & 0.000 & 0.950 & \textbf{0.980} & 0.000 & 0.810 & \textbf{0.900} & 0.000 & 0.870 & \textbf{0.940} & 1998 \\
Super Family 4 & 0.910 & 0.990 & \textbf{1.000} & 0.920 & \textbf{1.000} & \textbf{1.000} & 0.910 & 0.990 & \textbf{1.000} & 81559 \\
Super Family 5 & 0.870 & 0.950 & \textbf{0.970} & 0.880 & 0.940 & \textbf{0.970} & 0.870 & 0.950 & \textbf{0.970} & 40450 \\
Super Family 6 & 0.000 & 0.840 & \textbf{0.960} & 0.000 & 0.550 & \textbf{0.670} & 0.000 & 0.670 & \textbf{0.790} & 2877 \\
Super Family 7 & 0.000 & 0.830 & \textbf{0.940} & 0.000 & 0.680 & \textbf{0.910} & 0.000 & 0.750 & \textbf{0.920} & 3326 \\
Super Family 8 & 0.940 & 0.990 & \textbf{1.000} & \textbf{1.000} & 0.990 & \textbf{1.000} & 0.970 & 0.990 & \textbf{1.000} & 13267 \\
Super Family 9 & 0.000 & 0.980 & \textbf{1.000} & 0.000 & 0.910 & \textbf{1.000} & 0.000 & 0.940 & \textbf{1.000} & 52 \\
Super Family 10 & 0.000 & 0.000 & \textbf{0.910} & 0.000 & 0.000 & \textbf{0.830} & 0.000 & 0.000 & \textbf{0.870} & 11 \\
\hline
Micro Average & 0.896 & 0.977 & \textbf{0.990} & 0.919 & 0.979 & \textbf{0.992} & 0.903 & 0.980 & \textbf{0.988} & 28774 \T \\
Macro Average & 0.469 & 0.875 & \textbf{0.979} & 0.446 & 0.822 & \textbf{0.938} & 0.452 & 0.845 & \textbf{0.956} & 28774 \B \\
\end{tabular}
}
\end{center}
\end{table*}

\subsection{Model Interpretability}
We analyze the binary LSTM classifier in order to provide some intuition about the function of the various layers.  It is important to note that in the LSTM model, each layer in Fig. \ref{fig:model} is jointly optimized for the binary classification task.  Nevertheless, analyzing each layer independently does provide some intuition about the model's operation and performance.

\begin{figure}
\centering
\includegraphics[scale=0.45]{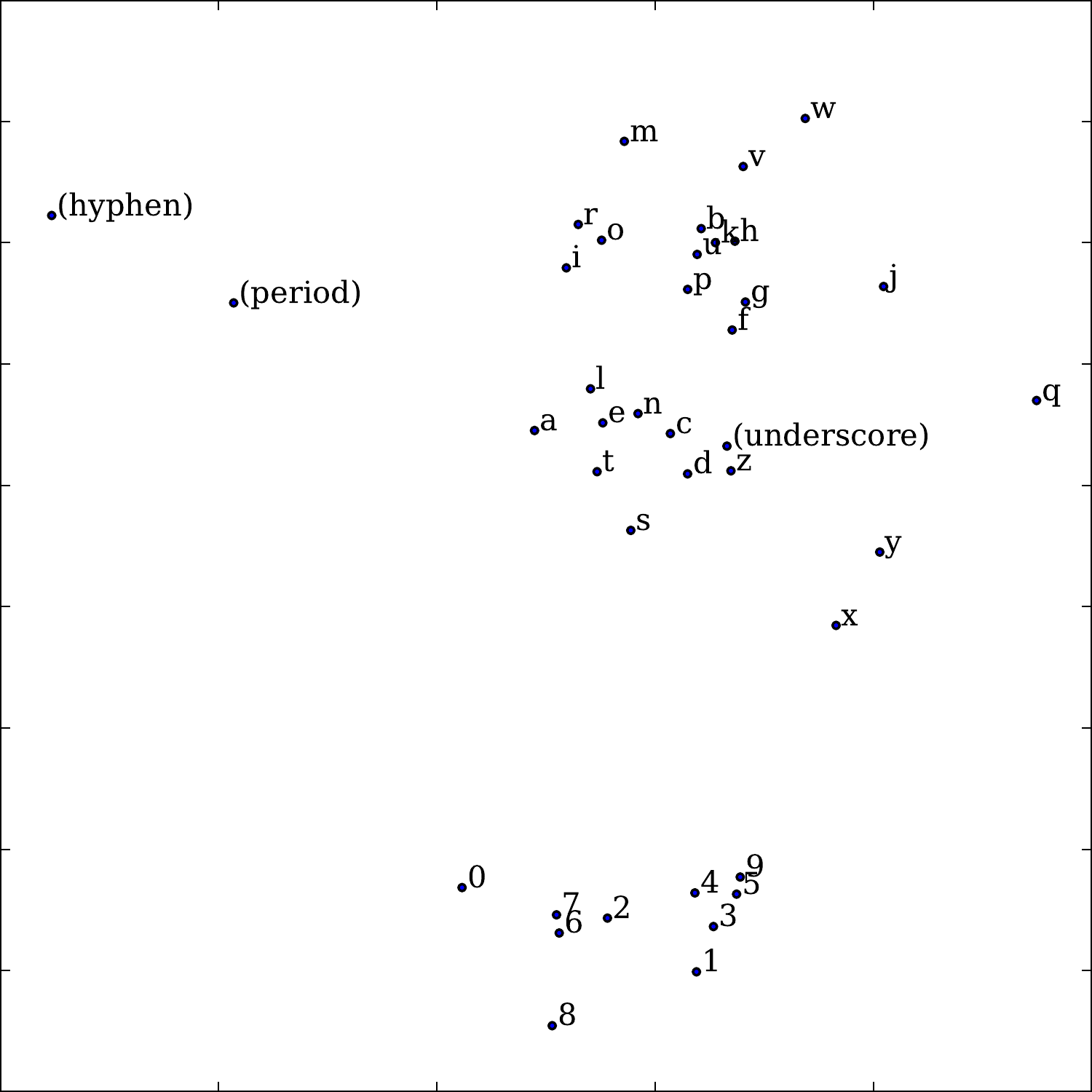}
\caption{Two-dimensional linear projection (PCA) of the embedded character vectors learned by the LSTM binary classifier.  Note that the model groups characters by similar effect on the LSTM layer's states and the subsequent model loss.}
\label{fig:char_vecs}
\end{figure}

The embedding layer in Fig. \ref{fig:model} learns a $128$-dimensional vector representation for each character in the set of valid domain characters.  A two-dimensional linear projection (via PCA) of the character embeddings is shown in Fig. \ref{fig:char_vecs}.  It is clear that the learned embedding consists of non-orthogonal vectors for each character.  This is in contrast to the orthonormal one-hot encoding of bigrams used in the logistic regression character bigram model.  The placement of vectors in the embedding space (and subsequently, the two-dimensional plot) relates to the similarity or interchangeability of characters for the DGA vs. non-DGA discrimination task.  For example, one would infer from the plot that replacing ``9'' with ``5'' would have much less effect on the score of the DGA classifier than would replacing ``9'' with ``w''.  The plot shows that there are obvious clusters of numeric digits and alphabetic characters (and underscore), while the less-common hyphen and period are fairly dissimilar to every other character.

Next, we investigate the \texttt{state} (or \texttt{memory}) of several LSTM cells in the second layer of the LSTM model in Fig. \ref{fig:model}.  The state of an LSTM cell has an initial value that is updated as each character of a domain is fed through the model.  It is a function of the current input (embedded character vector) and the previous emission of the LSTM cell.  In turn, the LSTM's emission is a function of the current state, current input, and previous emission.  In our model, the final emission (corresponding to the last character in the domain) from each of 128 LSTM cells is fed to the final logistic regression layer of the model to produce the DGA score.

\begin{figure*}
\begin{center}
\begin{tabular}{c}
\includegraphics[scale=0.5]{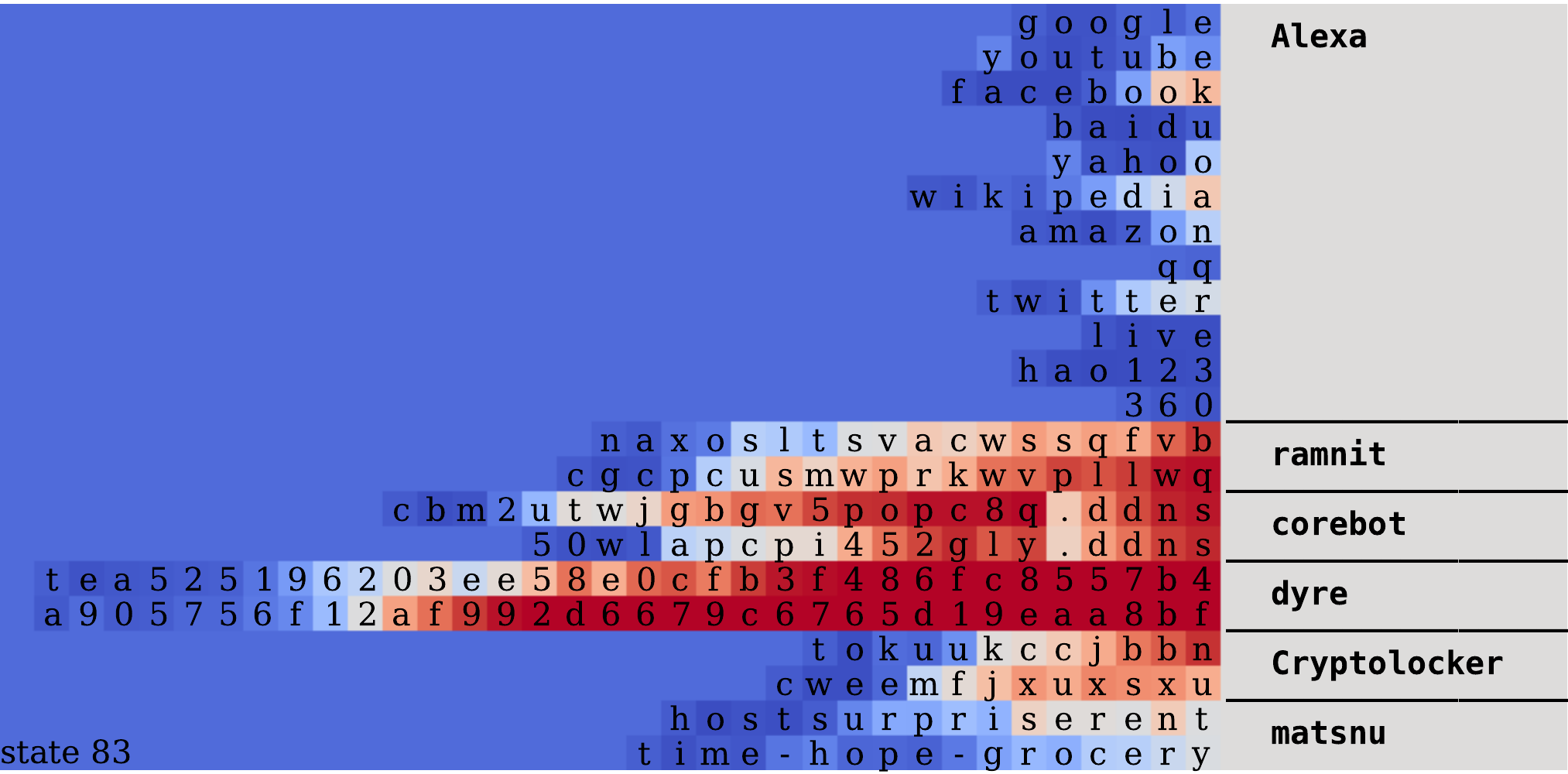} \\
(a) approximately tracks long domain names, with a soft reset on period and hypen\\
\includegraphics[scale=0.5]{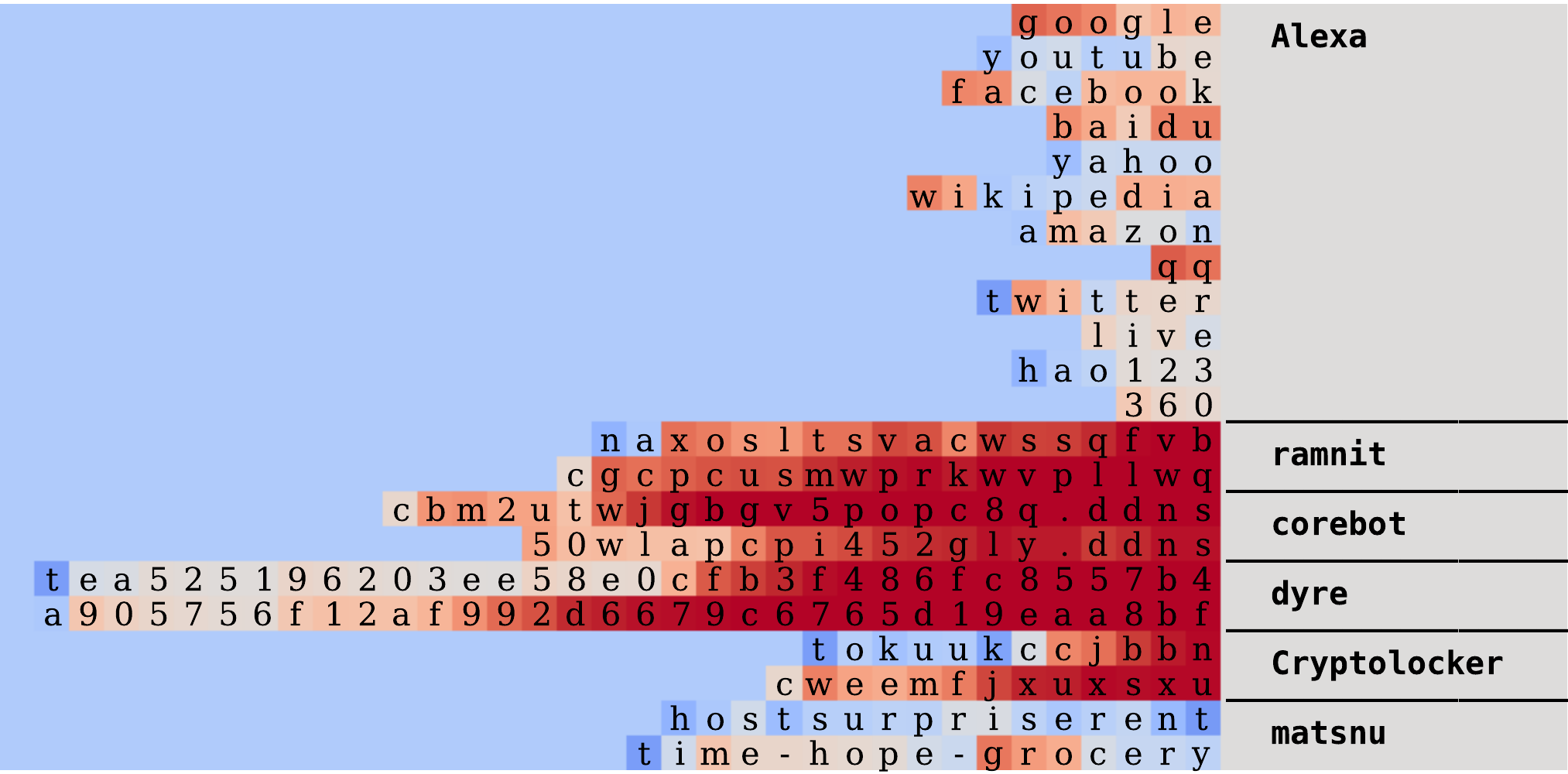} \\
(b) appears to track random alphanumeric sequences, as in \texttt{ramnit}, \texttt{corebot}, \texttt{dyre} and \texttt{Cryptolocker} \\
\includegraphics[scale=0.5]{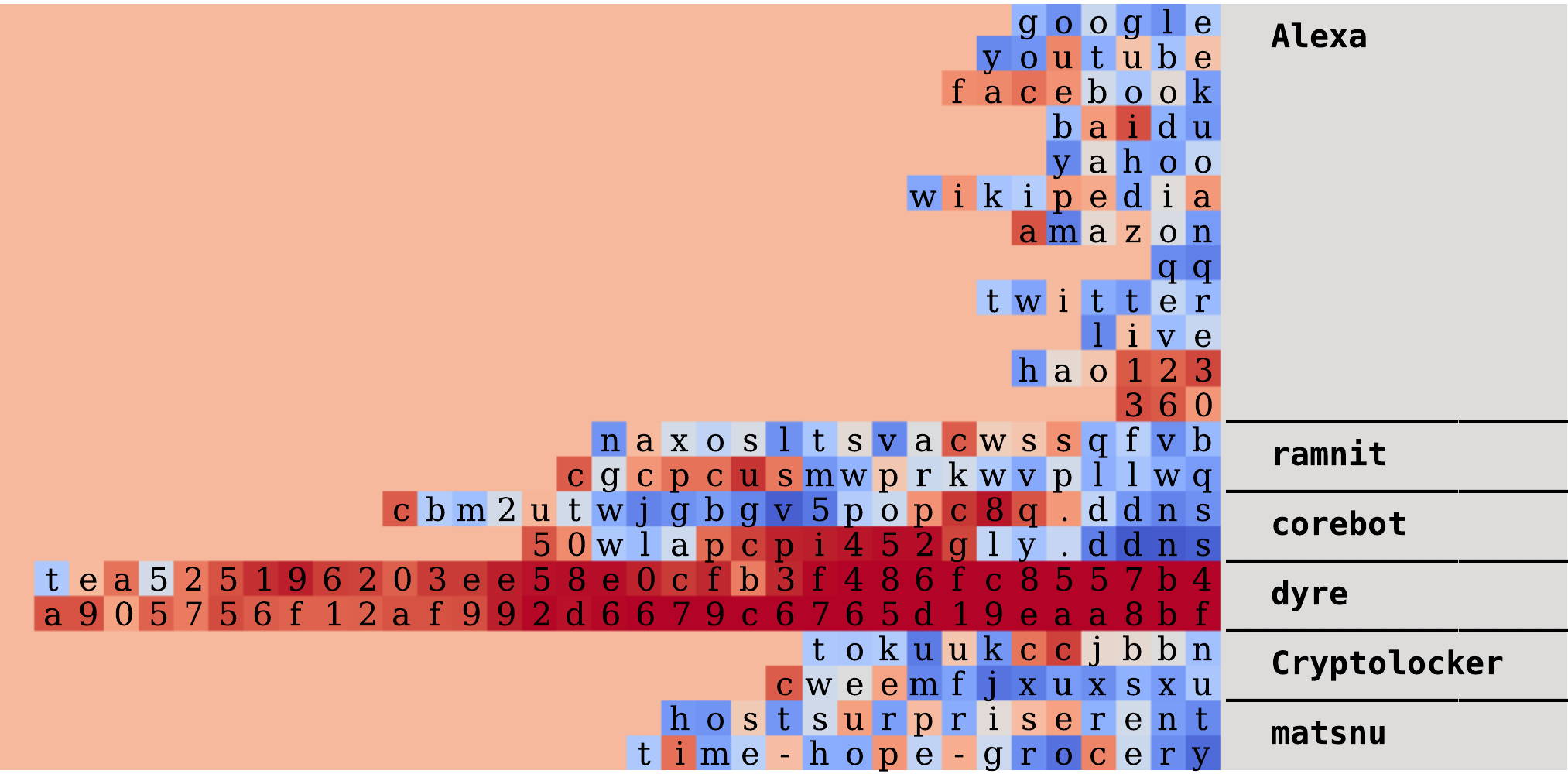} \\
(c) appears to track hexademical sequences, as in \texttt{dyre} \\
\includegraphics[scale=0.5]{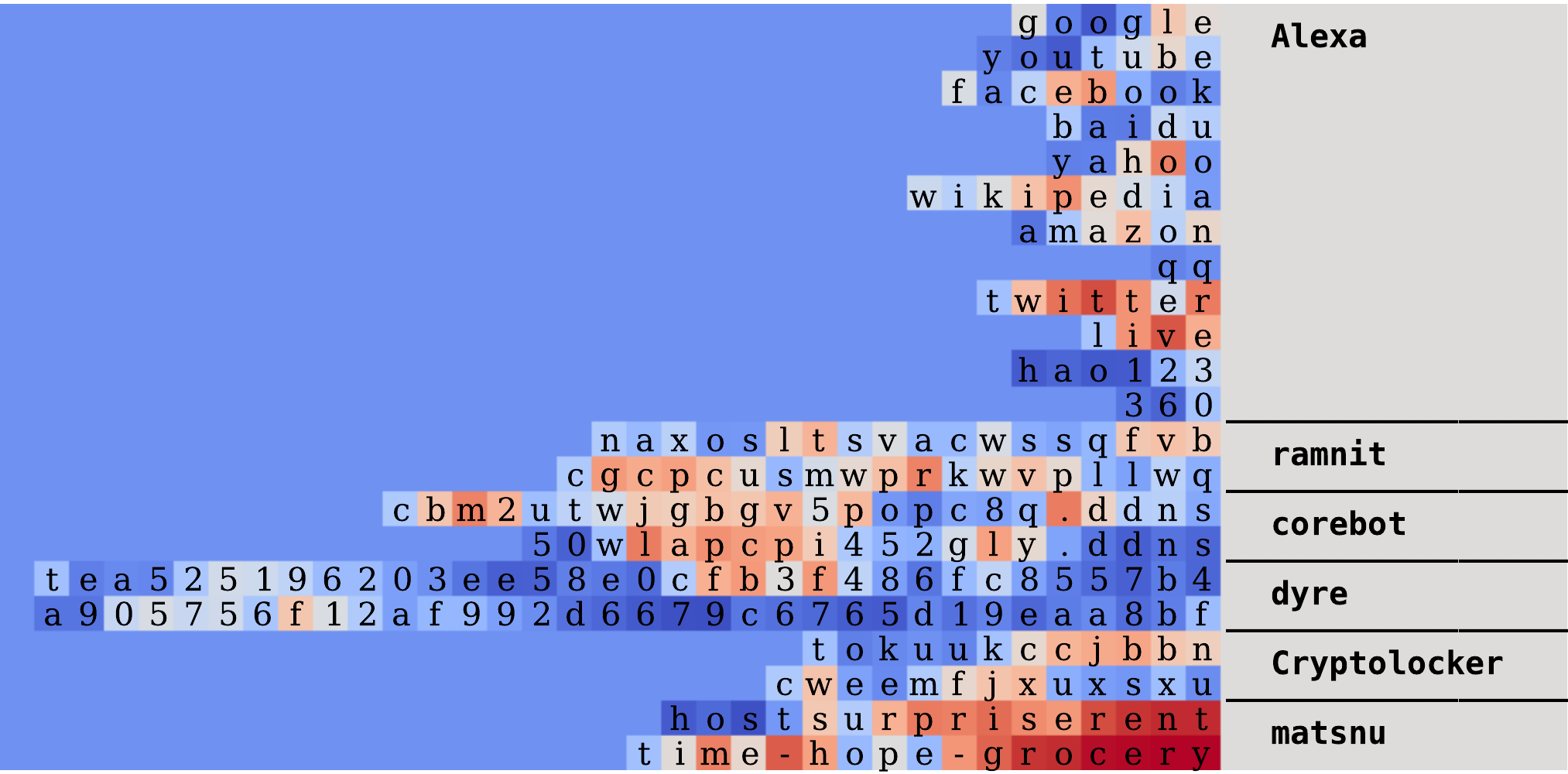} \\
(d) as in this example, it is difficult to ascribe an intuitive function of most states 
\end{tabular}
\end{center}
\caption{Examples of LSTM cell state values as domain characters are fed into the model. Color corresponds to the \textrm{tanh} of the state, and does not necessarily denote DGA or non-DGA.  Color preceeding a domain name denotes the cell's initial state. Our model correctly identifies DGA or non-DGA for all examples shown except for the final two \texttt{matsnu} examples.} \label{fig:states}
\end{figure*}

Each LSTM cell acts somewhat as an optimized feature extractor on the sequences of embedded character vectors produced from the previous embedding layer, and the cell's state provides an indication of what the cell is tracking.  Similar to \cite{karpathy2015visualizing}, Fig. \ref{fig:states} shows the \textrm{tanh} of a particular LSTM cell's state (called \textit{memory} in \cite{karpathy2015visualizing}) as it is updated character-by-character during a prediction task.  As shown in Fig. \ref{fig:states},  some states in our model have a tendency to track common characteristics of domain names in the dataset.  For example, Fig. \ref{fig:states}(a) shows a state that  seems to trend with domain name length, with soft resets on periods and hyphens.  The LSTM cell state depicted in Fig. \ref{fig:states}(b) appears to accumulate large values for long sequences of random alphanumeric characters.  The state in Fig. \ref{fig:states}(c) seems to accumulate value on sequences of hexadecimal characters, as is the predominant pattern in \texttt{dyre}.  Finally, Fig. \ref{fig:states}(d) depicts the most common scenario we encountered while inspecting states: it's generally very difficult to determine precisely what the state is tracking.  We note that our application of LSTMs for DGA classification does not yield quite as clearly the distinctive purpose of states as has been demonstrated for natural language models \cite{karpathy2015visualizing}.

\section{Conclusion}
\label{conclusion}
This paper presented an approach using LSTM networks to classify DGA generated domains.  LSTMs are advantageous over other techniques as they are featureless, using raw domain names as its input.  There is no need to manually create features that are difficult to maintain and can be rendered useless in an adversarial machine learning setting. In addition, an LSTM classifier can be run in real-time on single domains on standard commodity hardware making it trivial to deploy in virtually all security settings.  Experiments on publicly-available datasets showed that the LSTM classifier performed significantly better than other techniques (both real-time and retrospective), with the ability to classify 90\% of DGAs with a false positive rate of $10^{-4}$.  In addition, the LSTM classifier may be trivially modified for multiclass classification, which can provide context about the origin and intent of the domain-generating malware.

An in-depth analysis of results showed that the most difficult algorithms to classify are, intuitively, those that are modeled from a similar character distribution as domains in the Alexa top one million.  Some of these DGA families concatenate randomly selected words from (typically) English dictionaries.  However, the LSTM classifier was able to distinguish those DGA families when the amount of training examples were significant and the families were grouped together in super families.

We also provided an in-depth analysis of the functional interpretability of each layer in the LSTM DGA classifier.  Our analysis revealed that the model optimized vector embeddings for each character in a somewhat intuitive way, with distinct clusters for alphabetic and numeric digits.  Our analysis of the LSTM layer revealed the existence of LSTM cells that track a few somewhat interpretable features such as a hexadecimal  and random character sequences.  However, we found that most states did not provide clear interpretable evidence of function, in contrast to other applications of LSTMs, e.g., \cite{karpathy2015visualizing}.

Like all models, experiments show that our model is sensitive to class imbalance, which limits its ability to detect families with very little support in the training set (e.g., \texttt{matsnu}, \texttt{symmi} and \texttt{cryptowall}).  In the extreme case of zero training support, it was found that the LSTM model does not generalize well for detecting all families with very distinctive structure.  Manually-engineered features were able to detect some of those families that an LSTM classifier missed, and we hypothesize that this is directly a result of expert-tuned bias in the feature set that cannot be represented in the featureless LSTM model.  

All relevant source code and suggestions on deploying a real-world LSTM DGA classifier were provided by this paper.  In addition, we reference open datasets to create an equal classifier to that presented in this paper.  To the best of our knowledge, the presented system is by far the best performing DGA classification system as well as one of the easiest to deploy.



\bibliography{ref}	

\begin{thebibliography}{10}

\bibitem{robinson1994application}
A.~J. Robinson, ``An application of recurrent nets to phone probability
  estimation,'' {\em Neural Networks, IEEE Transactions on}, vol.~5, no.~2,
  pp.~298--305, 1994.

\bibitem{mikolov2010recurrent}
T.~Mikolov, M.~Karafi{\'a}t, L.~Burget, J.~Cernock{\`y}, and S.~Khudanpur,
  ``Recurrent neural network based language model.,'' in {\em INTERSPEECH},
  vol.~2, p.~3, 2010.

\bibitem{graves2012sequence}
A.~Graves, ``Sequence transduction with recurrent neural networks,'' {\em arXiv
  preprint arXiv:1211.3711}, 2012.

\bibitem{bengio2013advances}
Y.~Bengio, N.~Boulanger-Lewandowski, and R.~Pascanu, ``Advances in optimizing
  recurrent networks,'' in {\em Acoustics, Speech and Signal Processing
  (ICASSP), 2013 IEEE International Conference on}, pp.~8624--8628, IEEE, 2013.

\bibitem{hochreiter1997long}
S.~Hochreiter and J.~Schmidhuber, ``Long short-term memory,'' {\em Neural
  computation}, vol.~9, no.~8, pp.~1735--1780, 1997.

\bibitem{gers2000learning}
F.~A. Gers, J.~Schmidhuber, and F.~Cummins, ``Learning to forget: Continual
  prediction with lstm,'' {\em Neural computation}, vol.~12, no.~10,
  pp.~2451--2471, 2000.

\bibitem{gers2003learning}
F.~A. Gers, N.~N. Schraudolph, and J.~Schmidhuber, ``Learning precise timing
  with lstm recurrent networks,'' {\em The Journal of Machine Learning
  Research}, vol.~3, pp.~115--143, 2003.

\bibitem{srivastava2014dropout}
N.~Srivastava, G.~Hinton, A.~Krizhevsky, I.~Sutskever, and R.~Salakhutdinov,
  ``Dropout: A simple way to prevent neural networks from overfitting,'' {\em
  The Journal of Machine Learning Research}, vol.~15, no.~1, pp.~1929--1958,
  2014.

\bibitem{Alexa1M}
``Does alexa have a list of its top-ranked websites?.''
  \url{https://support.alexa.com/hc/en-us/articles/200449834-Does-Alexa-have-a-list-of-its-top-ranked-websites-}.
\newblock Accessed: 2016-04-06.

\bibitem{dgafeed}
``Bambenek consulting - master feeds.''
  \url{http://osint.bambenekconsulting.com/feeds/}.
\newblock Accessed: 2016-04-06.

\end{thebibliography}


\begin{thebibliography}{10}

\bibitem{kuhrer2014paint}
M.~K{\"u}hrer, C.~Rossow, and T.~Holz, ``Paint it black: Evaluating the
  effectiveness of malware blacklists,'' in {\em Research in Attacks,
  Intrusions and Defenses}, pp.~1--21, Springer, 2014.

\bibitem{antonakakis2012throw}
M.~Antonakakis, R.~Perdisci, Y.~Nadji, N.~Vasiloglou, S.~Abu-Nimeh, W.~Lee, and
  D.~Dagon, ``From throw-away traffic to bots: detecting the rise of
  {DGA}-based malware,'' in {\em P21st USENIX Security Symposium (USENIX
  Security 12)}, pp.~491--506, 2012.

\bibitem{yadav2010detecting}
S.~Yadav, A.~K.~K. Reddy, A.~Reddy, and S.~Ranjan, ``Detecting algorithmically
  generated malicious domain names,'' in {\em Proc. 10th ACM SIGCOMM conference
  on Internet measurement}, pp.~48--61, ACM, 2010.

\bibitem{yadav2012detecting}
S.~Yadav, A.~K.~K. Reddy, A.~N. Reddy, and S.~Ranjan, ``Detecting
  algorithmically generated domain-flux attacks with {DNS} traffic analysis,''
  {\em Networking, IEEE/ACM Transactions on}, vol.~20, no.~5, pp.~1663--1677,
  2012.

\bibitem{krishnan2013crossing}
S.~Krishnan, T.~Taylor, F.~Monrose, and J.~McHugh, ``Crossing the threshold:
  Detecting network malfeasance via sequential hypothesis testing,'' in {\em
  2013 43rd Annual IEEE/IFIP International Conference on Dependable Systems and
  Networks (DSN)}, pp.~1--12, IEEE, 2013.

\bibitem{chollet2016}
F.~Chollet, ``keras.'' \url{https://github.com/fchollet/keras}, 2016.

\bibitem{knysz2011good}
M.~Knysz, X.~Hu, and K.~G. Shin, ``Good guys vs. bot guise: Mimicry attacks
  against fast-flux detection systems,'' in {\em INFOCOM, 2011 Proceedings
  IEEE}, pp.~1844--1852, IEEE, 2011.

\bibitem{stone2011analysis}
B.~Stone-Gross, M.~Cova, B.~Gilbert, R.~Kemmerer, C.~Kruegel, and G.~Vigna,
  ``Analysis of a botnet takeover,'' {\em Security \& Privacy, IEEE}, vol.~9,
  no.~1, pp.~64--72, 2011.

\bibitem{ward2014cryptolocker}
M.~Ward, ``Cryptolocker victims to get files back for free,'' {\em BBC News,
  August}, vol.~6, 2014.

\bibitem{cryptolocker}
``A closer look at cyrptolocker's {DGA}.''
  \url{https://blog.fortinet.com/post/a-closer-look-at-cryptolocker-s-dga}.
\newblock Accessed: 2016-04-22.

\bibitem{hampton2015ransomware}
N.~Hampton and Z.~A. Baig, ``Ransomware: Emergence of the cyber-extortion
  menace,'' in {\em Australian Information Security Management Conference},
  2015.

\bibitem{cherepanov2013hesperbot}
A.~Cherepanov and R.~Lipovsky, ``Hesperbot-{A} new, advanced banking trojan in
  the wild,'' 2013.

\bibitem{w32ramnit}
Symantec, {\em W32.Ramnit analysis}.
\newblock 2015-02-24, Version 1.0.

\bibitem{suppoboxbh}
J.~Geffner, ``End-to-end analysis of a domain generating algorithm malware
  family.'' Black Hat USA 2013, 2013.

\bibitem{antonakakis2010building}
M.~Antonakakis, R.~Perdisci, D.~Dagon, W.~Lee, and N.~Feamster, ``Building a
  dynamic reputation system for {DNS}.,'' in {\em USENIX security symposium},
  pp.~273--290, 2010.

\bibitem{bilge2011exposure}
L.~Bilge, E.~Kirda, C.~Kruegel, and M.~Balduzzi, ``Exposure: Finding malicious
  domains using passive analaysis.,'' in {\em 18th Annual Network and
  Distributed System Security Symposium}, 2011.

\bibitem{bilge2014exposure}
L.~Bilge, S.~Sen, D.~Balzarotti, E.~Kirda, and C.~Kruegel, ``Exposure: a
  passive {DNS} analysis service to detect and report malicious domains,'' {\em
  ACM Transactions on Information and System Security (TISSEC)}, vol.~16,
  no.~4, p.~14, 2014.

\bibitem{schiavoni2014phoenix}
S.~Schiavoni, F.~Maggi, L.~Cavallaro, and S.~Zanero, ``Phoenix: {DGA}-based
  botnet tracking and intelligence,'' in {\em Detection of intrusions and
  malware, and vulnerability assessment}, pp.~192--211, Springer, 2014.

\bibitem{robinson1994application}
A.~J. Robinson, ``An application of recurrent nets to phone probability
  estimation,'' {\em Neural Networks, IEEE Transactions on}, vol.~5, no.~2,
  pp.~298--305, 1994.

\bibitem{mikolov2010recurrent}
T.~Mikolov, M.~Karafi{\'a}t, L.~Burget, J.~Cernock{\`y}, and S.~Khudanpur,
  ``Recurrent neural network based language model.,'' in {\em INTERSPEECH},
  vol.~2, p.~3, 2010.

\bibitem{graves2012sequence}
A.~Graves, ``Sequence transduction with recurrent neural networks,'' {\em arXiv
  preprint arXiv:1211.3711}, 2012.

\bibitem{bengio2013advances}
Y.~Bengio, N.~Boulanger-Lewandowski, and R.~Pascanu, ``Advances in optimizing
  recurrent networks,'' in {\em Acoustics, Speech and Signal Processing
  (ICASSP), 2013 IEEE International Conference on}, pp.~8624--8628, IEEE, 2013.

\bibitem{hochreiter1997long}
S.~Hochreiter and J.~Schmidhuber, ``Long short-term memory,'' {\em Neural
  computation}, vol.~9, no.~8, pp.~1735--1780, 1997.

\bibitem{gers2000learning}
F.~A. Gers, J.~Schmidhuber, and F.~Cummins, ``Learning to forget: Continual
  prediction with {LSTM},'' {\em Neural computation}, vol.~12, no.~10,
  pp.~2451--2471, 2000.

\bibitem{gers2003learning}
F.~A. Gers, N.~N. Schraudolph, and J.~Schmidhuber, ``Learning precise timing
  with {LSTM} recurrent networks,'' {\em J. Machine Learning Research}, vol.~3,
  pp.~115--143, 2003.

\bibitem{Alexa1M}
``Does {A}lexa have a list of its top-ranked websites?\killpunct.''
  \url{https://support.alexa.com/hc/en-us/articles/200449834-Does-Alexa-have-a-list-of-its-top-ranked-websites-}.
\newblock Accessed: 2016-04-06.

\bibitem{dgafeed}
``Bambenek consulting - master feeds.''
  \url{http://osint.bambenekconsulting.com/feeds/}.
\newblock Accessed: 2016-04-06.

\bibitem{porras2009foray}
P.~A. Porras, H.~Sa{\"\i}di, and V.~Yegneswaran, ``A foray into conficker's
  logic and rendezvous points.,'' in {\em LEET}, 2009.

\bibitem{andriesse2013highly}
D.~Andriesse, C.~Rossow, B.~Stone-Gross, D.~Plohmann, and H.~Bos, ``Highly
  resilient peer-to-peer botnets are here: An analysis of gameover zeus,'' in
  {\em Malicious and Unwanted Software:" The Americas"(MALWARE), 2013 8th
  International Conference on}, pp.~116--123, IEEE, 2013.

\bibitem{kraken}
P.~Royal, ``On the kraken and bobax botnets.''
  \url{https://www.damballa.com/downloads/r_pubs/Kraken_Response.pdf}, 2008.

\bibitem{stone2009your}
B.~Stone-Gross, M.~Cova, L.~Cavallaro, B.~Gilbert, M.~Szydlowski, R.~Kemmerer,
  C.~Kruegel, and G.~Vigna, ``Your botnet is my botnet: analysis of a botnet
  takeover,'' in {\em Proceedings of the 16th ACM conference on Computer and
  communications security}, pp.~635--647, ACM, 2009.

\bibitem{karpathy2015visualizing}
A.~Karpathy, J.~Johnson, and F.-F. Li, ``Visualizing and understanding
  recurrent networks,'' in {\em to appear in Proceedings of the International
  Conference on Learning Representations}, 2016.
\newblock arXiv preprint arXiv:1506.02078.

\end{thebibliography}
\bibliographystyle{ieeetr}

\end{document}